\documentclass[aps,prb,reprint,letterpaper,superscriptaddress,amsmath,amssymb,floatfix]{revtex4-2}

\usepackage{graphicx}
\usepackage{bm}
\usepackage{xcolor}
\usepackage{url}
\usepackage[colorlinks=true,allcolors=blue]{hyperref}

\newcommand{\ang}{\,\text{\AA}}
\newcommand{\angi}{\,\text{\AA}^{-1}}
\newcommand{\EF}{E_F}
\newcommand{\eh}{\,e^2/h}
\newcommand{\ehA}{\,e^2h^{-1}\text{\AA}^2}
\newcommand{\NbSe}{NbSe$_2$}
\newcommand{\slight}{\sigma^{\rm light}_{xy}}
\newcommand{\sex}{\sigma_{\rm ex}}
\newcommand{\Heff}{H_{\rm eff}}
\newcommand{\MVcm}{\,MV/cm}
\newcommand{\T}{\mathcal{T}}

\begin{document}

\title{Optical Hall valve and light-induced Berry curvature dipole in monolayer Ising metals}

\author{Yusuf Wicaksono}
\email{Contact author: WICAKSONO.Yusuf@nims.go.jp}
\author{Toshikaze Kariyado}
\email{Contact author: KARIYADO.Toshikaze@nims.go.jp}
\affiliation{International Center for Materials Nanoarchitectonics (WPI-MANA), National Institute for Materials Science, 1-1 Namiki, Tsukuba, Ibaraki 305-0044, Japan}

\date{\today}

\begin{abstract}
Light polarization offers contactless symmetry control of Hall responses. Using a Wannier-based finite-frequency Floquet theory with micromotion corrections, benchmarked against Floquet diagonalization, we study monolayer 1H-NbSe$_2$ driven below its dominant interband absorption onset. Circular polarization controls the anomalous Hall conductivity, whereas linear polarization induces a Berry curvature dipole by making the van Hove saddles inequivalent; at a photon energy of 0.2~eV and an rms field of 0.4~MV/cm, its hole-doped peak reaches $5.6\times10^{-2}$ and $3.4\times10^{-2}$~\AA{} at electron temperatures of 50 and 300~K, respectively. With out-of-plane proximity exchange, the helicity sets the sign of the Hall conductivity for exchange up to about 30~meV at leading order in the field: an optical Hall valve. These predictions assume Fermi--Dirac occupations of the dressed bands and neglect the charge-density wave.
\end{abstract}

\maketitle

% ---------------------------------------------------------------------
% Introduction
% ---------------------------------------------------------------------
\textit{Introduction.} Hall responses are dictated by symmetry. The anomalous Hall conductivity $\sigma_{xy}$ requires broken time-reversal symmetry $\T$, whereas in $\T$-invariant metals a nonlinear Hall response arises from the Berry curvature dipole (BCD) $\bm{D}$, an in-plane pseudovector forbidden by threefold rotation $C_3$~\cite{Sodemann2015,Ma2019,Kang2019,Du2021,Ortix2021}. In trigonal monolayers such as the 1H transition-metal dichalcogenides, a BCD therefore requires an external symmetry breaker: uniaxial strain~\cite{You2018,Son2019} or, as we showed for 1H-NbX$_2$, layer-selective magnetic proximity~\cite{Wicaksono2026}. These knobs are static, and magnetic proximity additionally requires control of the magnetic configuration.

Light is a contactless, reversible and ultrafast alternative. Circularly polarized light breaks $\T$, lifts valley degeneracies~\cite{Sie2015} and induces Floquet topological phases and an anomalous Hall effect~\cite{Oka2009,Kitagawa2011,Lindner2011,Sato2019,McIver2020,Zhan2023}. Floquet--Bloch bands have been resolved in photoemission~\cite{Wang2013,Ito2023,Zhou2023}, and Floquet control of Berry curvature dipoles and related nonlinear photocurrents has been proposed for tilted Dirac bands, warped topological-insulator surfaces, where linear light lifts the $C_3$ constraint, and driven transition-metal dichalcogenides~\cite{Qin2024,PriorTI2026,PriorMoSe2,Prior1TMoS2}. Two questions remain open for real materials. First, are light-induced Hall responses competitive with those of static symmetry breakers off resonance, where interband absorption is suppressed and heating is limited? Second, do the high-frequency expansions that underlie most Floquet design~\cite{Oka2019,Rudner2020,Eckardt2015,Bukov2015,Mikami2016} remain reliable there? For a metal whose bands span several eV, photon energies of a fraction of an eV lie outside the high-frequency regime and make converged full-Floquet calculations demanding~\cite{Vogl2020,RodriguezVega2021}.

Here we answer both questions for monolayer 1H-\NbSe, an Ising metal~\cite{Xi2016} and a common building block of magnetic van der Waals heterostructures~\cite{Kezilebieke2020}. We combine a first-principles Wannier model, which permits the dense $k$ sampling that the BCD requires, with a finite-frequency Floquet effective Hamiltonian that includes the micromotion correction to the Berry curvature and is benchmarked against Floquet diagonalization, and we target the Fermi-surface BCD of a metal and the competition between helicity and magnetic proximity. We find that (i) the Stokes parameters of the light select the Hall response: helicity controls $\sigma_{xy}$ and the polarization axis controls $\bm D$; (ii) mid-infrared linear light at $\hbar\omega=0.2$~eV and 0.4\MVcm{} makes the van Hove saddles at the three M points inequivalent and generates a hole-doped peak BCD of $5.6\times10^{-2}$ and $3.4\times10^{-2}\ang$ at electron temperatures of 50 and 300~K, respectively, above the peak BCD that 30~meV of layer-selective exchange produces in the same model~\cite{Wicaksono2026}; and (iii) with a weak out-of-plane proximity exchange, the helicity sets the sign of $\sigma_{xy}$ at the Fermi level for exchange up to $\Delta^*\approx6$ and 30~meV at 0.2 and 0.4\MVcm{} at leading order in the field: an optical Hall valve. These fields are comparable to those of mid-infrared Floquet experiments~\cite{Wang2013,McIver2020}, and we map where the theory is controlled and heating is moderate. In this regime the light-induced Hall conductivity has the opposite sign to its value in the high-frequency regime ($\hbar\omega=10$~eV), and the BCD is several times larger.

% ---------------------------------------------------------------------
% Model and method
% ---------------------------------------------------------------------
\begin{figure}[t]
\includegraphics[width=\columnwidth]{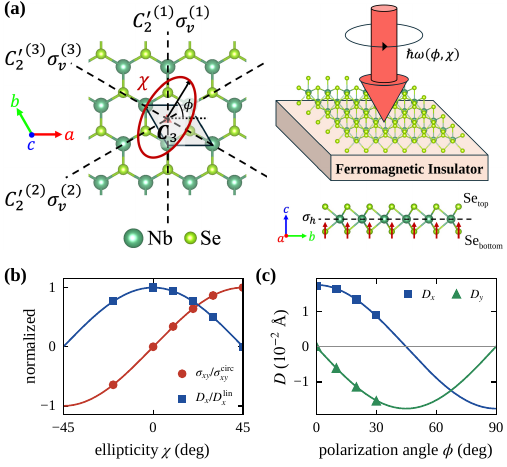}
\caption{Polarization selection rules. (a) Left: top view of 1H-\NbSe{} with the $C_3$ axis (triangle) and the mirror planes $\sigma_v^{(i)}$, which contain the axes $C_2^{\prime(i)}$ (dashed); red: polarization ellipse of Eq.~\eqref{eq:drive} (angle $\phi$ from $x\parallel\bm a$, ellipticity $\chi$). Right: light at normal incidence (here circular, $\eta=+1$) on 1H-\NbSe{} on a ferromagnetic insulator; side view: moments induced on the bottom Se layer ($\bm m\parallel\hat{\bm z}$, exchange $\Delta$) break the mirror $\sigma_h$. (b) $\sigma_{xy}$ (circles) and $D_x$ (squares) versus $\chi$, normalized by projection onto the circular- and linear-light curves over the full $\mu$ window; lines: $\sin2\chi$ and $\cos2\chi$. (c) $D_x$ (squares) and $D_y$ (triangles) versus $\phi$ for linear light at the BCD peak ($\mu-\EF=-0.215$~eV); lines: $\beta e^{-2i\phi}$ with $\beta$ from the data at $0$--$30^\circ$, the symmetry-reduced range. $\Delta=0$ in (b,c); $\hbar\omega=0.2$~eV, $\alpha=0.01\angi$ ($E_0=0.2$\MVcm), $T=50$~K.}
\label{fig:rules}
\end{figure}

\textit{Model and drive.} We use the 22-band spinor Wannier model (Nb $d$, Se $p$) of Ref.~\cite{Wicaksono2026}, obtained from fully relativistic density-functional theory~\cite{Giannozzi2009,Giannozzi2017,Perdew2008,Marzari1997,Souza2001,Pizzi2020}. Its Hamiltonian, position matrix and Wannier centres are symmetrized under the $C_{3v}$ subgroup of the crystal point group before any dressing, and $\EF$ is the DFT Fermi level~\cite[Secs.~S1.2 and S1.3]{SMnote}. Normally incident light enters through the in-plane vector potential
\begin{equation}
\bm A(t)=\sqrt2A_0\left[\cos\chi\,\bm e_1\cos\omega t+\sin\chi\,\bm e_2\sin\omega t\right],
\label{eq:drive}
\end{equation}
with $\bm e_1=(\cos\phi,\sin\phi,0)$, $\bm e_2=\hat{\bm z}\times\bm e_1$, polarization angle $\phi$ and ellipticity $\chi$ [$\chi=0$: linear; $\chi=\pm45^\circ$: circular, with $\eta=+1$ ($-1$) for counterclockwise (clockwise) rotation of the field seen from $+\hat{\bm z}$; Fig.~\ref{fig:rules}(a)]. The cycle-averaged intensity is independent of $\chi$. We quote the drive strength $\alpha=eA_0/\hbar$, related to the root-mean-square field by $E_0\,[\mathrm{MV/cm}]=100\,\alpha\,[\text{\AA}^{-1}]\,\hbar\omega\,[\mathrm{eV}]$. Light couples through Peierls phases on the full intersite vectors plus the dipole coupling $e\bm E(t)\cdot(\bm r-\bm\tau)$ of the residual position matrix~\cite{Ventura2017}, with $\bm\tau$ the Wannier centres; the bands are then independent of the orbital-centre convention to within $10^{-5}$~eV~\cite[Sec.~S2.1]{SMnote}.

\textit{Finite-frequency Floquet theory.} For $\hbar\omega$ far below the bandwidth, Floquet replicas fold densely: the high-frequency expansion fails, and our Brillouin-zone-integrated full-Floquet calculations do not converge under the tested truncations and occupation prescriptions, although converged diagonalization at selected nonresonant $k$ points provides a benchmark~\cite[Secs.~S5.2 and S6.4]{SMnote}. Away from resonances, however, the central Floquet sector can be decoupled by a Schrieffer--Wolff (Floquet--L\"owdin) transformation~\cite{Bukov2016,Mikami2016,Eckardt2015}. To second order, in the eigenbasis of the time-averaged Hamiltonian $H^{(0)}$ (energies $\varepsilon_n$),
\begin{multline}
[\Heff]_{nm}=\varepsilon_n\delta_{nm}+\tfrac12\sum_{p\neq0}\sum_l H^{(-p)}_{nl}H^{(p)}_{lm}\\
\times\left[g(\varepsilon_n-\varepsilon_l-p\hbar\omega)+g(\varepsilon_m-\varepsilon_l-p\hbar\omega)\right],
\label{eq:heff}
\end{multline}
where $H^{(p)}$ are the Fourier components of the driven Hamiltonian and $g(x)=x/(x^2+\gamma^2)$ regularizes small denominators with $\gamma=\hbar/\tau=26$~meV ($\tau=25$~fs); this phenomenological regularization, motivated by the scattering time, is not a self-consistent self-energy~\cite[Sec.~S6.2]{SMnote}. For $\hbar\omega\to\infty$, Eq.~\eqref{eq:heff} reduces to the familiar $H^{(0)}+\sum_{p>0}[H^{(p)},H^{(-p)}]/(p\hbar\omega)$~\cite{Bukov2015,Eckardt2015,Mikami2016}. The Berry curvature of the Floquet states contains, besides the curvature of $\Heff$ evaluated by Wannier interpolation~\cite{Wang2006}, a micromotion term: the curl of the additional Berry connection generated by the transformation~\cite{Weinberg2017}, which we evaluate including position-matrix elements and as a trace over degenerate or nearly degenerate band groups~\cite[Sec.~S6.3]{SMnote}. At 10~eV the integrated light-induced $\sigma_{xy}$ agrees with full Floquet to three decimal places; at 0.2~eV and $\alpha=0.01\angi$, the light-induced Berry curvature at sampled nonresonant $k$ points agrees within 2--4\% (linear light) and 4--10\% (circular light with exchange), with errors scaling as $\alpha^4$~\cite[Sec.~S6.4]{SMnote}. The helicity-odd curvature is more sensitive, because the $\Heff$ and micromotion contributions nearly cancel: its median absolute error is 0.36 times its median magnitude at $\alpha=0.01\angi$ and 1.4 times at $0.02\angi$. Its $\alpha^2$ coefficient, integrated over the zone on three grids, agrees within 0.7\% with exact Floquet states in the weak-field limit, but its higher-order correction is not controlled; we therefore report the valve threshold at leading order~\cite[Sec.~S6.4]{SMnote}.

\textit{Responses.} We compute~\cite{Xiao2010,Sodemann2015}
\begin{equation}
\sigma_{xy}=-\frac{e^2}{\hbar}\sum_n\int_{\bm k}f_n\Omega_n,\qquad
D_a=\sum_n\int_{\bm k}f_n\,\partial_{k_a}\Omega_n,
\label{eq:resp}
\end{equation}
with $\int_{\bm k}\equiv\int d^2k/(2\pi)^2$, $\Omega_n$ the Floquet Berry curvature and $f_n$ the occupations. To leading order, the light-induced Hall current is third order in drive and probe fields~\cite{Murotani2026}, the BCD-mediated one fourth order; the Floquet form gives both as dressed-band geometry~\cite{Morimoto2016} with explicit occupations~\cite[Sec.~S6.7]{SMnote}. We assume Fermi--Dirac occupations of the dressed bands at the electron temperature ($T=50$~K unless stated, as in Ref.~\cite{Wicaksono2026}) and fixed chemical potential; at 50~K, fixing the density instead changes the light-induced Hall conductivity and $\bm D$ at $\EF$ by less than 0.3\%~\cite[Sec.~S5.4]{SMnote}. This approximation is motivated off resonance~\cite{Seetharam2015,Dehghani2014} and fails near resonance~\cite{Cao2024}; here suppressed interband absorption supports but does not establish it: without exchange, the horizontal mirror forbids in-plane transitions between the Ising-split Nb branches (layer-selective exchange activates them only weakly), the dominant interband absorption sets in at 0.60~eV, and the Nb--Se-$p$ near-resonance at $\gtrsim0.35$~eV connects two occupied bands and is Pauli blocked. A kinetic calculation for a specific pulse and environment would be needed to determine the distribution; a wide-band bath, for instance, would fill the photon replicas~\cite[Secs.~S5.2 and S5.3]{SMnote}. Below the dominant onset we approximate the absorbed power by its Drude part, which enters through an estimated electron temperature (see below). Grids of $1200^2$ ($\sigma_{xy}$) and $2400^2$ ($\bm D$) $k$ points are converged to within 1\% (rms) under grid refinement, and the band grouping changes the BCD peak by at most 1.7\%~\cite[Secs.~S6.3 and S6.5]{SMnote}. As a practical diagnostic of perturbative validity, we monitor the largest element of the regularized Schrieffer--Wolff generator, $|H^{(p)}_{ln}\,g(\varepsilon_n-\varepsilon_l-p\hbar\omega)|$, for states within $\EF\pm0.3$~eV; resonances, where this element is bounded by $|H^{(p)}_{ln}|/2\gamma$ and can be small, are assessed separately~\cite[Sec.~S6.6]{SMnote}. It stays below 0.2 at 0.2~eV for $E_0\le0.40$\MVcm{} with circular light and 0.30\MVcm{} with linear light. For the BCD at 0.4\MVcm{} it reaches 0.27; there the median absolute error of the light-induced curvature against exact Floquet states at sampled $k$ points is 6\% of its median magnitude~\cite[Sec.~S6.6]{SMnote}. Results at stronger fields indicate trends.

% ---------------------------------------------------------------------
% Selection rules (Fig. 1)
% ---------------------------------------------------------------------
\textit{Polarization selection rules.} Without light or exchange, $\T$ forces $\sigma_{xy}=0$ and $C_3$ forces $\bm D=0$, both obeyed to numerical precision by the symmetrized model~\cite[Sec.~S1.3]{SMnote}. Light lowers the symmetry in a way fixed by its Stokes parameters. The helicity $S_3=\sin2\chi$ is $\T$-odd and $C_3$-invariant, like $\sigma_{xy}$. The linear components $S_1+iS_2=\cos2\chi\,e^{2i\phi}$ are $\T$-even and define a headless in-plane axis, while $\bm D$ is a $\T$-even in-plane pseudovector~\cite{Sodemann2015}. Under $C_3$, $D_\pm=D_x\pm iD_y$ acquires $e^{\pm2\pi i/3}$ while $\phi\to\phi+2\pi/3$, so $D_+$ may contain only harmonics $e^{in\phi}$ with even $n\equiv1\pmod3$; at second order in the field only $n=-2$ survives. Hence, to this order and without exchange,
\begin{equation}
\sigma_{xy}\propto\sin2\chi,\qquad D_x+iD_y=\beta\cos2\chi\,e^{-2i\phi},
\label{eq:rules}
\end{equation}
where $\beta\propto\alpha^2$ depends on $\mu$, $\hbar\omega$ and $T$ and is real because, for linear light at $\phi=0$, the mirror $\sigma_v^{(1)}$ ($x\to-x$) is preserved and forbids $D_y$. To leading order, rotating the polarization axis by $\delta\phi$ rotates $\bm D$ by $-2\delta\phi$. Figure~\ref{fig:rules} confirms Eq.~\eqref{eq:rules} at 0.2~eV and $\alpha=0.01\angi$: $\sigma_{xy}$ follows $\sin2\chi$ within 0.03\% and $\bm D$ follows $\cos2\chi$ within 1.2--2.5\%, with higher harmonics below 2\% of $\beta$; circular light gives $|\bm D|\le6\times10^{-5}\ang$; and, without exchange, $\bm D(\chi)=\bm D(-\chi)$ holds exactly. With exchange $\Delta$ added, $\T$ further implies $\sigma_{xy}(-\Delta,-\eta)=-\sigma_{xy}(\Delta,\eta)$ and $\bm D(-\Delta,-\eta)=\bm D(\Delta,\eta)$, both verified numerically~\cite[Sec.~S3]{SMnote}.

% ---------------------------------------------------------------------
% BCD (Fig. 2)
% ---------------------------------------------------------------------
\begin{figure}[t]
\includegraphics[width=\columnwidth]{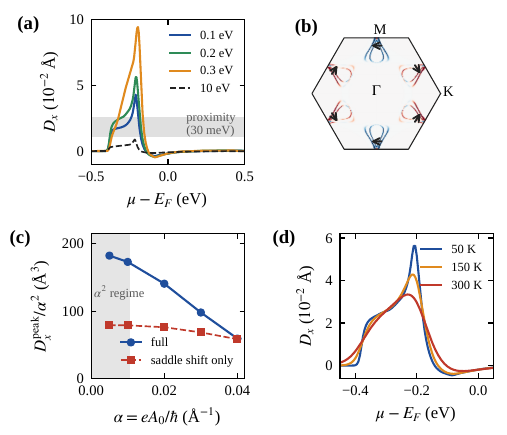}
\caption{Light-induced BCD for linear light ($\phi=0$, $T=50$~K unless stated). (a) $D_x$ versus chemical potential at $\alpha=0.02\angi$ for $\hbar\omega=0.1$, 0.2 and 0.3~eV, and the high-frequency effective-Hamiltonian result at 10~eV (computed with WannierBerri~\cite{Tsirkin2021}, dashed); at 0.1 and 0.3~eV ($E_0=0.2$ and 0.6\MVcm) the drive exceeds the linear-light validity bound (trends only), and at 0.2~eV it is checked against exact Floquet states. Shaded: peak BCD for 30~meV layer-selective exchange (orthogonal and canted textures) in the same model, $T=50$~K~\cite{Wicaksono2026}. (b) Undriven BCD kernel $v_x\Omega(-f')$, $v_x=\partial\varepsilon/\partial k_x$, at the BCD-peak energy, $\mu-\EF=-0.215$~eV, just below the M saddle ($T=50$~K; red/blue: positive/negative); arrows: dipoles integrated over the parts of $\pm0.25\angi$ boxes around the six M points that lie inside the first zone; they cancel by $C_3$. (c) Peak $D_x/\alpha^2$ versus $\alpha$ at 0.2~eV: full calculation (circles) and energy-shift term of Eq.~\eqref{eq:saddle} (squares); shaded: $\alpha^2$ regime. (d) $D_x$ at 50, 150 and 300~K (0.2~eV, $\alpha=0.02\angi$).}
\label{fig:bcd}
\end{figure}

\textit{Light-induced BCD.} Linear light at 0.2~eV produces a BCD peaked about 0.2~eV below $\EF$ [Fig.~\ref{fig:bcd}(a)]: $D_x=1.732\times10^{-2}\ang$ at $\alpha=0.01\angi$ ($E_0=0.2$\MVcm, $\mu-\EF\approx-0.215$~eV) and $5.63\times10^{-2}\ang$ at $\alpha=0.02\angi$ ($E_0=0.4$\MVcm, $-0.205$~eV). Like the proximity peaks~\cite{Wicaksono2026}, these peaks lie away from $\EF$ and require doping: within the rigid-band approximation, about $6.8\times10^{14}$ holes per cm$^2$ (0.69 per Nb), a density at which structural and self-consistent electronic changes could modify the peak. At $\EF$ (0.2~eV, 0.4\MVcm), $D_x=-1.8\times10^{-3}\ang$~\cite[Sec.~S8.4]{SMnote}. At fixed $\alpha=0.02\angi$, the peak grows from $4.29\times10^{-2}\ang$ at 0.1~eV to $7.07\times10^{-2}\ang$ at 0.25~eV, several times the high-frequency value ($0.94\times10^{-2}\ang$ at 10~eV, full Floquet). Mid-infrared light thus exceeds the peak BCD that 30~meV of layer-selective exchange produces in the same model, $(1.1$--$2.6)\times10^{-2}\ang$ for orthogonal and canted textures~\cite{Wicaksono2026} [shaded band in Fig.~\ref{fig:bcd}(a)], without a magnet and with an orientation set by $\phi$; the comparison is between separately located peaks, not at equal doping.

The origin is a hidden dipole at the van Hove saddles of the Nb $d$ band, at $E_{\rm vH}-\EF=-0.2065$~eV [Fig.~\ref{fig:bcd}(b)]. Regions around the three inequivalent M points contribute large dipoles, $D_x\approx-0.24\ang$ for the M point on the $k_y$ axis and $+0.12\ang$ for each of the other two ($\pm0.25\angi$ boxes), which cancel by $C_3$. Linear light breaks the equivalence of the three saddles: at $\alpha=0.01\angi$ and $\phi=0$ it raises the saddle at the M point perpendicular to the field by 2.95~meV and the other two by only 0.43~meV, and it dresses their dipoles unequally. Writing the dipole of region $j$ as $D_+^{(j)}=d(\mu-\delta E_j)[1+\kappa_0+\kappa_2\cos2(\phi-\theta_j)]e^{i(\theta_j-\pi/2)}$, with $\theta_j$ the direction of M$_j$ ($\theta_1=\pi/2$ on the $k_y$ axis) and saddle shifts $\delta E_j=\delta_0+\delta_2\cos2(\phi-\theta_j)$, $C_3$ symmetry gives, to leading order in the shifts,
\begin{equation}
D_x+iD_y=\tfrac32\left[\delta_2\,\partial_\mu d(\mu)-\kappa_2\,d(\mu)\right]e^{-2i\phi},
\label{eq:saddle}
\end{equation}
where $d(\mu)$ is the hidden dipole of a single saddle. With $d=-0.233\ang$, $\partial_\mu d=-3.05$~\AA/eV, $\delta_2=-1.69$~meV and $\kappa_2=+0.024$ at $\mu-\EF=-0.215$~eV, the energy-shift term, evaluated with the finite shifts, contributes $0.79\times10^{-2}\ang$ and the dipole-dressing term $0.84\times10^{-2}\ang$; their sum, $1.63\times10^{-2}\ang$, is 94\% of the full result. Equation~\eqref{eq:saddle} accounts for the peak position near the saddle energy, the $-2\phi$ rotation, and the decrease with temperature [$(5.63,\,4.28,\,3.35)\times10^{-2}\ang$ at 50, 150 and 300~K, respectively; Fig.~\ref{fig:bcd}(d)], which smears the sharp derivative $\partial_\mu d$. The energy-shift term alone reproduces the peak position and its drift with $\alpha$ [Fig.~\ref{fig:bcd}(c)]. The response becomes subquadratic once the light-induced saddle splitting ($\approx2.5$~meV at $\alpha=0.01\angi$, growing as $\alpha^2$) exceeds $k_BT$: $D/\alpha^2$ falls from 182~\AA$^3$ at $\alpha=0.005\angi$ to 141~\AA$^3$ at $0.02\angi$ [Fig.~\ref{fig:bcd}(c)], mostly through the dressing term, since the shift term alone falls more slowly. The BCD changes by less than 5\% when $\gamma$ is halved or doubled or the dipole coupling is removed~\cite[Sec.~S10]{SMnote}. In 1H-NbS$_2$ the same mechanism gives a peak at its own M saddle, nine times smaller~\cite[Secs.~S8.4 and S11]{SMnote}.

% ---------------------------------------------------------------------
% Hall valve (Fig. 3)
% ---------------------------------------------------------------------
\begin{figure}[t]
\includegraphics[width=\columnwidth]{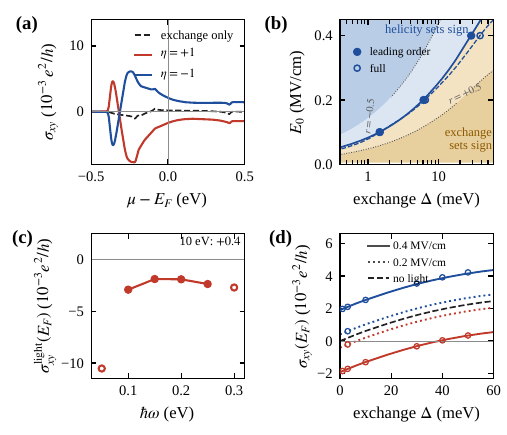}
\caption{Optical Hall valve at $\hbar\omega=0.2$~eV and $T=50$~K (unless stated) with out-of-plane exchange $\Delta$ on the bottom Se layer. (a) $\sigma_{xy}$ for $\Delta=3$~meV without light (dashed) and with circular light of helicity $\eta=\pm1$ at $\alpha=0.02\angi$ ($E_0=0.4$\MVcm). (b) Helicity contrast $r=\sigma_{xy}(\EF;\eta=+1)/\sigma_{xy}(\EF;\eta=-1)$ at leading order in the field: the helicity sets the sign for $r<0$ (blue) and the exchange for $r>0$ (beige); dotted: $r=\pm0.5$; solid line and filled circles: $\Delta^*$ ($r=0$) and its values at $E_0=0.1$, 0.2 and 0.4\MVcm; dashed line and open circles: the same with the full $\slight(\EF)$. (c) $\slight(\EF)$ versus $\hbar\omega$ at $\alpha=0.02\angi$; open symbols: $E_0$ about twice the conservative validity bound (trends only). The value at 10~eV is $+0.4\times10^{-3}\eh$. (d) $\sigma_{xy}(\EF)$ for $\eta=+1$ (red) and $-1$ (blue) versus $\Delta$: lines, $\sex+\eta\slight$ at 0.4 and 0.2\MVcm{} and without light; circles, full calculations with exchange.}
\label{fig:valve}
\end{figure}

\textit{Optical Hall valve.} We define the light-induced Hall conductivity as the helicity-odd part without exchange, $\slight\equiv[\sigma_{xy}(\eta=+1)-\sigma_{xy}(\eta=-1)]_{\Delta=0}/2$; we approximate it by the helicity-odd response at $\Delta=3$~meV, which agrees within 0.01\% at $\EF$ for $\hbar\omega=0.2$~eV and $\alpha=0.02\angi$~\cite[Sec.~S9.1]{SMnote}. Circular light alone gives $\slight(\EF)=-1.909\times10^{-3}\eh$ at 0.2~eV and $\alpha=0.02\angi$, and between $-2.9\times10^{-3}$ and $-1.9\times10^{-3}\eh$ from 0.1 to 0.25~eV [Fig.~\ref{fig:valve}(c)]. At 10~eV, in the high-frequency regime, it has the opposite sign, $+4\times10^{-4}\eh$ (full Floquet, same $\alpha$, i.e., a 50 times stronger field); qualitative $k$-resolved maps locate the reversal in the K/K$'$ pockets. The reversal is a micromotion effect: without the micromotion term, the FFEH gives $+1.1\times10^{-2}\eh$ at 0.2~eV~\cite[Secs.~S6.4 and S9.4]{SMnote}. Even the sign of the light-induced Hall response of a metal driven below its interband scale therefore cannot be inferred from the high-frequency limit.

We now add an out-of-plane exchange $\Delta s_z$ ($s_z$: Pauli matrix) on the $p$ orbitals of the bottom Se layer, as induced by an adjacent ferromagnetic insulator~\cite{Wicaksono2026,Kezilebieke2020}. Exchange alone yields $\sex(\EF)$ sublinear in $\Delta$, from $6.63\times10^{-5}\eh$ at 1~meV to $1.595\times10^{-3}\eh$ at 30~meV. Guided by the symmetry $\sigma_{xy}(-\Delta,-\eta)=-\sigma_{xy}(\Delta,\eta)$, we test the decomposition $\sigma_{xy}(\Delta,\eta)=\sex(\Delta)+\eta\,\slight(\alpha)+\delta\sigma(\Delta,\eta)$; the cross term is small: at $\EF$ the helicity-odd part stays between $-1.940\times10^{-3}$ and $-1.909\times10^{-3}\eh$ for $\Delta=1$--50~meV, within 1.6\% of the light-only value, and the helicity-even part reproduces $\sex$ within $3.5\times10^{-5}\eh$. Light and exchange are thus additive at $\EF$ over the tested range [Fig.~\ref{fig:valve}(d): the sum reproduces all full calculations within $7\times10^{-5}\eh$]: the helicity reverses the light-induced part, while the magnetization, and with it the helicity-even part $\sex$, stays fixed. Figure~\ref{fig:valve}(a) shows the consequence at $\Delta=3$~meV: the light-induced part (peak $6.9\times10^{-3}\eh$ at $\mu-\EF=-0.235$~eV) dominates the exchange contribution, and the two helicities give opposite signs of $\sigma_{xy}$ at $\EF$, as over most of the window shown~\cite[Sec.~S9.1]{SMnote}.

\begin{figure}[t]
\includegraphics[width=\columnwidth]{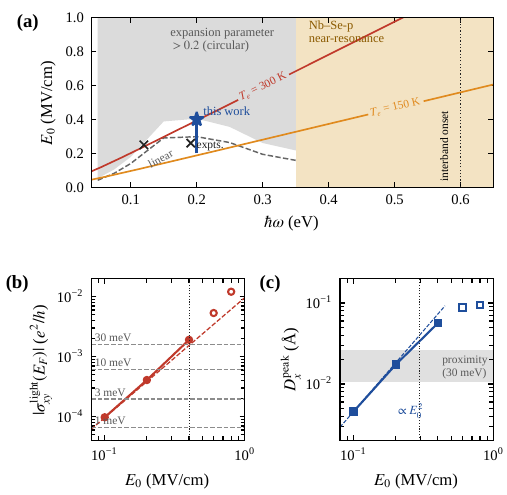}
\caption{Experimental window. (a) Gray: the Schrieffer--Wolff expansion parameter exceeds 0.2 for some state within $\EF\pm0.3$~eV (circular light); dashed: the same bound for linear light; beige: Nb--Se-$p$ near-resonance; dotted: dominant interband absorption onset. Lines: electron temperature $T_e$ after a 100~fs pulse starting from 50~K ($\tau=25$~fs, $\sigma_{xx}=5$~mS, mass-enhanced heat capacity). Star and bar: operating points of Figs.~\ref{fig:bcd} and \ref{fig:valve}; crosses: field amplitudes at the sample in Floquet experiments~\cite{Wang2013,McIver2020}. (b) $|\slight(\EF)|$ and (c) peak $D_x$ versus $E_0$ at 0.2~eV. Dashed lines: $\sex(\EF)$ for $\Delta=1$, 3, 10 and 30~meV (gray) and leading order $|c_2|\alpha^2$ (red) in (b), $E_0^2$ scaling in (c); shaded in (c): proximity-induced BCD for 30~meV exchange~\cite{Wicaksono2026}. Dotted: validity bound at 0.2~eV for circular (b) and linear (c) light; open symbols lie beyond it, and the BCD at 0.4\MVcm{} is checked against exact Floquet states~\cite[Sec.~S6.6]{SMnote}.}
\label{fig:window}
\end{figure}

Within the additive approximation, the valve operates while $|\slight(\EF)|>|\sex(\EF)|$, i.e., below the first crossover $\Delta^*$. Away from resonances our scheme is correct through second order in the field as $\gamma\to0$ (halving $\gamma$ changes $\slight(\EF)$ by 0.1\%), so we estimate the threshold from $\slight(\EF)\simeq c_2\alpha^2$ with $c_2=-3.85\ehA$. The higher-order part retained by the scheme, 6\% at $\alpha=0.01\angi$ and 19\% at $0.02\angi$ [Fig.~\ref{fig:window}(b)], is not controlled and provides no error bound~\cite[Sec.~S6.4]{SMnote}. At leading order [Fig.~\ref{fig:valve}(b)], $\Delta^*=1.5$, 6.1 and 29~meV for $E_0=0.1$, 0.2 and 0.4\MVcm, interpolated from exchange-only calculations up to 120~meV (39~meV with the full approximate expression at 0.4\MVcm). Of $c_2$, 27\% is a Fermi-surface population term, in which helicity-odd band shifts act on the equilibrium Berry curvature; with occupations inherited from the undriven bands it is absent, the sign is unchanged, and $\Delta^*=20$~meV at 0.4\MVcm~\cite[Sec.~S5.4]{SMnote}. $\Delta^*$ grows faster than $E_0^2$ because $\sex$ is sublinear in $\Delta$. Above $\Delta^*$ the exchange sets the sign, but in the additive estimate (tested up to 50~meV) the helicity still changes $\sigma_{xy}(\EF)$ by at least a factor of 3.8 at 0.4\MVcm{} [Figs.~\ref{fig:valve}(b) and \ref{fig:valve}(d)]~\cite[Sec.~S9.2]{SMnote}. The light-induced conductivity depends weakly on temperature, $\slight(\EF)=(-1.91,\,-1.93,\,-1.99,\,-2.13)\times10^{-3}\eh$ at 50, 150, 300 and 540~K, respectively, and changes by $-6\%$ without the dipole coupling~\cite[Secs.~S9.3 and S10]{SMnote}. Other light--exchange combinations give no comparable effect~\cite[Sec.~S4.3]{SMnote}.

The light-induced part is similar across the three compounds, $c_2=-3.1$, $-3.85$ and $-3.7\ehA$ for NbS$_2$, \NbSe{} and NbTe$_2$, but the exchange response is not: NbS$_2$ also works as a valve, with $\Delta^*\approx25$~meV at 0.4\MVcm, whereas the much larger exchange response of NbTe$_2$ near $\EF$ ($-4.7\times10^{-4}\eh$ per meV at $\Delta=3$~meV) limits its valve to $\Delta^*\approx1$~meV at $\alpha=0.01\angi$~\cite[Sec.~S11]{SMnote}.

% ---------------------------------------------------------------------
% Working window and detection (Fig. 4)
% ---------------------------------------------------------------------

\textit{Experimental window.} Figure~\ref{fig:window}(a) maps where these predictions hold, and Fig.~\ref{fig:window}(b,c) gives the field dependence of both responses at 0.2~eV. Below the dominant interband absorption onset of the model (0.60~eV; 0.42~eV at the hole-doped BCD peak; $T=0$ occupations), virtual Nb--Se-$p$ transitions approach resonance above $\approx0.35$~eV, and the expansion parameter limits $E_0$ to 0.18, 0.39, 0.40, 0.36 and 0.26\MVcm{} at $\hbar\omega=0.1$, 0.15, 0.2, 0.25 and 0.3~eV (0.05\MVcm{} at 0.05~eV; 25\% lower for linear light), comparable to the field amplitudes at the sample, about 0.25\MVcm{} at 0.12--0.19~eV, in Floquet experiments~\cite{Wang2013,McIver2020} (for circular light, $E_0$ equals the field amplitude). The frequency scans at fixed $\alpha=0.02\angi$ [Figs.~\ref{fig:bcd}(a) and \ref{fig:valve}(c)] include points beyond this conservative diagnostic. For circular light, $E_0$ exceeds the bound by 12\% and 41\% at 0.1 and 0.25~eV (95th-percentile bounds 0.19 and 1.09\MVcm) and about twofold at 0.05 and 0.3~eV; these two, like the linear-light results at 0.1 and 0.3~eV, indicate trends only. Intraband absorption heats the electrons: for $\tau=25$~fs and $\alpha=0.02\angi$, sheet conductances of 1--5~mS absorb 2.7--14~MW/cm$^2$ and a 100~fs pulse heats the lattice by 2--23~K. For 5~mS, the electron temperature rises from 50~K to $\approx160$~K at $\alpha=0.01\angi$ and $\approx300$~K at $\alpha=0.02\angi$ for a mass-enhanced electronic heat capacity, or to $\approx260$ and $\approx520$~K for the bare band value, nearly independently of $\hbar\omega$ at fixed $\alpha$ for $\omega\tau\gg1$; at 300~K the light-induced Hall conductivity at $\EF$ changes by 4\% and the BCD peak drops by 41\% [Fig.~\ref{fig:bcd}(d)], at 550~K by 12\% and 55\% (fixed $\mu$), and the leading-order valve window does not appreciably narrow~\cite[Secs.~S9.3 and S12.1]{SMnote}. These estimates call for short pulses; continuous-wave driving would require a separate heat-flow analysis. Pulses of several optical cycles, 100--200~fs at 0.2~eV and thus several times longer than $\tau$, allow Floquet dressing to build up~\cite{Ito2023}, while the occupations remain an assumption.

Taking as an illustrative value the sheet conductance $\sigma_{xx}\approx5$~mS measured near $T_c$ in monolayer \NbSe~\cite{Xi2016}, and the leading-order $\slight$, the valve at 0.4\MVcm{} corresponds to a Hall angle of $1.2\times10^{-5}$ and a transient Hall current of $\approx6$~nA at 0.1~V bias, only $\approx0.004$ electrons per 100~fs pulse. For a 5~$\mu$m device at 0.1~V, the estimated transient BCD-induced Hall current is 1--10~pA at $\EF$ and 40--310~pA at the hole-doped peak ($\tau=20$--150~fs); its quadratic bias dependence permits second-harmonic isolation in principle, but a time-averaged measurement is reduced by the duty cycle $t_pf_{\rm rep}$~\cite[Sec.~S12.3]{SMnote}. Detecting these transient signals is the main experimental challenge; it calls for ultrafast on-chip detection~\cite{McIver2020} or circular-dichroism time-resolved photoemission, which maps Floquet Berry curvature~\cite{Kim2026} and could probe the Berry-curvature redistribution near the saddles. For the ideal circularly driven valve geometry, normal incidence and preserved $C_3$ forbid bias-independent in-plane photocurrents odd in $\eta$; device asymmetry can relax this, so helicity and bias reversal should be combined. For the BCD, bias-independent photocurrents do not contribute at the second harmonic of the bias.

Monolayer \NbSe{} develops a $3\times3$ charge-density wave (CDW), reported below $T_{\rm CDW}\approx145$~K~\cite{Xi2015} with a gap of about 4~meV at $\EF$~\cite{Ugeda2016}, which we neglect, as in Ref.~\cite{Wicaksono2026}. Our predictions therefore describe the unreconstructed normal state and apply most directly to samples held above $T_{\rm CDW}$. From a 150~K base, the same pulses heat the electrons to $\approx210$ and $\approx335$~K ($\approx300$ and $\approx540$~K for the bare heat capacity); the light-induced Hall conductivity at $\EF$ then changes by at most 12\%, and the BCD peak (maximum over $\mu$) falls from $4.3\times10^{-2}\ang$ at 150~K to $(3.2,\,2.6)\times10^{-2}\ang$ at 335 and 540~K. Below $T_{\rm CDW}$, the small lattice heating does not determine the transient CDW dynamics: a reconstructed-band, nonequilibrium analysis would be needed, and a reconstruction near M could substantially modify the BCD peak.

% ---------------------------------------------------------------------
% Conclusion
% ---------------------------------------------------------------------
\textit{Conclusion.} Light polarization provides complementary, optically switchable symmetry breakers for the Hall responses of a trigonal metal: helicity for $\sigma_{xy}$ and the polarization axis for $\bm D$. In monolayer \NbSe, mid-infrared fields below 0.5\MVcm{} produce a hole-doped peak BCD exceeding the peak from 30~meV magnetic proximity and, together with weak exchange, an optical Hall valve whose sign follows the helicity for exchange up to about 30~meV at leading order. Both effects depend on the finite photon energy through the detunings in Eq.~\eqref{eq:heff} and the micromotion correction, both absent from the leading high-frequency effective Hamiltonian: compared with the mid-infrared, the response at 10~eV has a several times smaller BCD and the opposite sign of the light-induced $\sigma_{xy}$. Finite-frequency Floquet theory is therefore essential for light-controlled responses of metals driven below their interband scale; our Wannier-based implementation applies to other trigonal metals and heterostructures.

\textit{Acknowledgments.} The calculations in this study were performed on the Numerical Materials Simulator at NIMS. This work was supported by JSPS KAKENHI Grant No.~JP24K06968 from MEXT, Japan. Y.W. thanks Ryo Kitaura and Daichi Kozawa for fruitful discussions.

\textit{Data availability.} The data and codes supporting this study are available from the authors upon reasonable request.

\bibliography{refs_arxiv}

% =====================================================================
%                     Supplemental Material
% =====================================================================
\clearpage
\onecolumngrid
% title block and abstract as produced by revtex for the standalone supplement.tex (author marks as revtex sets them;
% abstract: 400 pt, \small); positions agree with supplement.pdf to about 0.1 pt
\begin{center}
\vspace*{-1.175em}
{\large\bfseries Supplemental Material for ``Optical Hall valve and light-induced Berry curvature dipole in monolayer Ising metals''\par}
\vspace{1.25em}
{\spaceskip=4.91pt\relax Yusuf Wicaksono\textsuperscript{1}\textsuperscript{,\,}\textsuperscript{\hyperlink{SMfn1}{$\ast$}} and Toshikaze Kariyado\textsuperscript{1}\textsuperscript{,\,}\textsuperscript{\hyperlink{SMfn2}{$\dagger$}}\par}
\vspace{0.35em}
{\small\itshape $^{1}$International Center for Materials Nanoarchitectonics (WPI-MANA),\\
National Institute for Materials Science, 1-1 Namiki, Tsukuba, Ibaraki 305-0044, Japan\par}
\vspace{0em}
{\small (Dated: \today)\par}
\vspace{0.75em}
\begin{minipage}{400pt}
\small\setlength{\parindent}{1em}\finalhyphendemerits=5000\relax  % LaTeX default (center sets 0), as in the revtex abstract
\indent This Supplemental Material documents the model, the light--matter coupling, the Floquet methods used (the
high-frequency effective Hamiltonian, exact diagonalization of the truncated Floquet Hamiltonian with the
occupation models tested, and the finite-frequency Floquet effective Hamiltonian (FFEH) with micromotion), their
validation and convergence, and the additional data behind every statement of the main text that cites it.
Occupation models are discussed separately from the Hamiltonian approximations.
The material is organized as follows.
Section~S1 describes the fully relativistic Wannier model of monolayer \NbSe{}, its symmetrization, the
proximity exchange and the models of NbS$_2$ and NbTe$_2$.
Section~S2 gives the light--matter coupling, the Floquet blocks and the response functions, together with
consistency checks and benchmarks.
Section~S3 derives the polarization selection rules for the Hall conductivity and the Berry curvature dipole
(BCD), without and with exchange.
Section~S4 treats the high-frequency limit, for light alone and combined with exchange.
Section~S5 describes the full Floquet steady state and the occupations: the method, its results and limits, the
occupation model used in the main text and the sensitivity of the results to the occupations.
Section~S6 develops the FFEH: the effective Hamiltonian, its regularization, the micromotion correction to the
Berry curvature, the validation against full Floquet calculations, including an integrated weak-field benchmark
against exact Floquet states, the convergence, the validity map and the relation to perturbative nonlinear-response
theory.
Section~S7 verifies the selection rules in the mid-infrared.
Section~S8 analyzes the mechanism of the light-induced BCD, from the hidden dipoles at the M saddles to a
saddle-point model, and gives the BCD at the Fermi level and in NbS$_2$ and a comparison with proximity-induced
BCD.
Section~S9 defines the optical Hall valve and gives its additivity, sign-control crossover and field,
frequency and temperature dependence, as well as the origin of its sign change relative to the high-frequency
limit.
Section~S10 summarizes the sensitivity of the main results to the broadening, the dipole coupling, the electron
temperature and numerical parameters, and Section~S11 gives the results for NbS$_2$ and NbTe$_2$.
Finally, Section~S12 estimates Drude heating and the working window, discusses the charge-density wave and gives
device estimates and competing signals.\par
\end{minipage}
\end{center}
\vspace{0.952em}
\twocolumngrid

% S-numbering of figures, tables, equations and pages (sections are numbered by hand: \section*{Sn. ...});
% unique hyperref anchors
\setcounter{figure}{0}\setcounter{table}{0}\setcounter{equation}{0}
\renewcommand{\theHfigure}{S\arabic{figure}}
\renewcommand{\theHtable}{S\arabic{table}}
\renewcommand{\theHequation}{S\arabic{equation}}
\setcounter{page}{1}
\renewcommand{\thepage}{S\arabic{page}}

\setcounter{figure}{0}
\renewcommand{\thefigure}{S\arabic{figure}}

\setcounter{table}{0}
\renewcommand{\thetable}{S\arabic{table}}

\setcounter{equation}{0}
\renewcommand{\theequation}{S\arabic{equation}}

% =====================================================================
\section*{S1. Wannier model and conventions}
% =====================================================================
Throughout this Supplemental Material, equation and figure numbers without the prefix S refer to the main text. Unless stated otherwise: 1H-\NbSe{}, $T=50$~K, $\EF=-2.4083$~eV (DFT), $\hbar\omega=0.2$~eV, broadening $\gamma=\hbar/\tau=26$~meV, FFEH, $\sigma_{xy}$ on a $1200^2$ and $\bm D$ on a $2400^2$ grid, and $\mu-\EF\in[-0.5,0.5]$~eV in steps of 5~meV (1~meV with WannierBerri).

% e-mail notes of the title block
\begin{figure}[!b]
\vspace*{12.0pt}\vspace{-0.6em}\noindent\rule{0.2\columnwidth}{0.4pt}\\[0.2em]
{\footnotesize\raggedright
\hypertarget{SMfn1}{}\hspace*{1.33pt}$^{*}$\hspace{1.41pt}\,\href{mailto:WICAKSONO.Yusuf@nims.go.jp}{Contact author: WICAKSONO.Yusuf@nims.go.jp}\\[0.35pt]
\hypertarget{SMfn2}{}\hspace*{1.72pt}$^{\dagger}$\hspace{1.41pt}\,\href{mailto:KARIYADO.Toshikaze@nims.go.jp}{Contact author: KARIYADO.Toshikaze@nims.go.jp}\par}
\vspace{-0.21pt}
\end{figure}

\begin{figure*}[t]
\includegraphics[width=\textwidth]{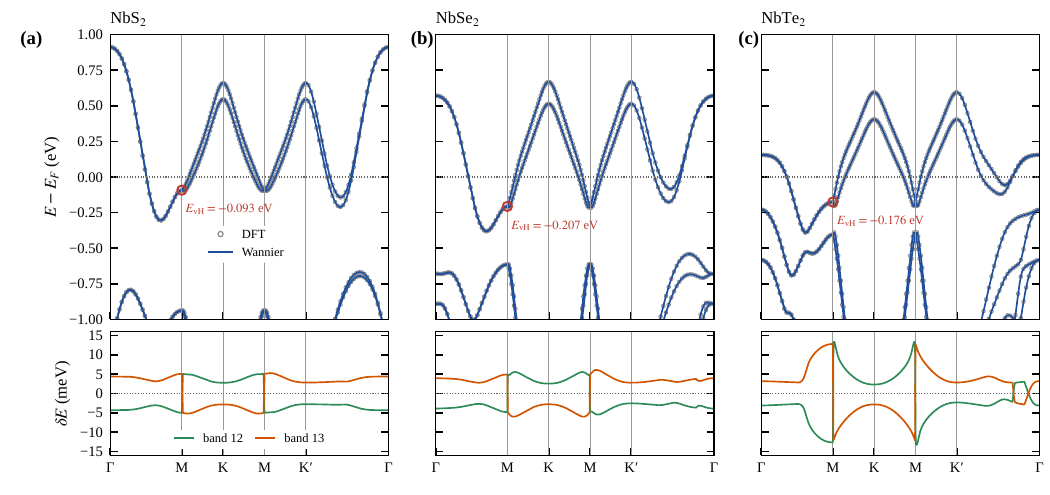}
\caption{Bands of 1H-NbS$_2$ (a), 1H-\NbSe{} (b) and 1H-NbTe$_2$ (c) along $\Gamma$--M--K--M--K$'$--$\Gamma$, relative to the DFT Fermi levels: fully relativistic DFT (symbols) and the 22-orbital Wannier model (lines); circles mark the M saddle of the Nb pair. Lower row: exchange shift $\delta E=E(\Delta)-E(0)$ of the Nb pair (bands 12 and 13 in energy order) for $\Delta=30$~meV, $\hat{\bm m}\parallel\hat{\bm z}$ on the bottom chalcogen layer. The steep steps at M are continuous on a finer path: the energy-ordered bands exchange their dominant spin where the spin--orbit order of the pair reverses at this time-reversal-invariant point; in NbTe$_2$, the pair also anticrosses a Te-$p$ band between K$'$ and $\Gamma$. Wannier bands of the unsymmetrized models of Ref.~\cite{S:Wicaksono2026}; symmetrization changes them by $\le0.4$~meV.}
\label{S:fig:bands}
\end{figure*}

\subsection*{S1.1. First-principles input}
The electronic structure is that of Ref.~\cite{S:Wicaksono2026}: fully relativistic density-functional theory in \textsc{Quantum ESPRESSO}~\cite{S:Giannozzi2009,S:Giannozzi2017} with the PBEsol functional~\cite{S:Perdew2008}, fully relativistic projector-augmented-wave pseudopotentials (pslibrary), a 90~Ry cutoff, Marzari--Vanderbilt smearing of 0.01~Ry and a $60\times60$ non-self-consistent $k$ mesh; $a=3.4225$~\AA{} and $c=30$~\AA{} (vacuum along $z$). The lattice vectors are $\bm a_1=a(1,0,0)$ and $\bm a_2=a(-\tfrac12,\tfrac{\sqrt3}{2},0)$, so that $x\parallel\bm a$ is a zigzag direction and the Nb--Se bonds project onto the armchair directions [main-text Fig.~1(a)].

\subsection*{S1.2. Maximally localized Wannier functions}
We use 22 spinor maximally localized Wannier functions~\cite{S:Marzari1997,S:Souza2001,S:Pizzi2020} (13 electrons): orbitals 1--10 are Nb $d$ ($d_{z^2}$, $d_{xz}$, $d_{yz}$, $d_{x^2-y^2}$, $d_{xy}$), 11--16 top-Se $p$ and 17--22 bottom-Se $p$ ($p_z$, $p_x$, $p_y$), with spin-interleaved spinors and $z$ quantization (checked against the Wannier90 projections). In energy order, bands 0--11 are Se $p$, bands 12--13 form the Nb pair ($d_{z^2}$ with $d_{x^2-y^2}/d_{xy}$) that crosses $\EF$, and bands 14--21 are the upper Nb $d$ manifold, separated in \NbSe{} by gaps of $\ge0.35$ and 1.43~eV. The Fermi level is the DFT value, $\EF=-2.4083$~eV, rather than the 13-electron filling of the Wannier model ($-2.4142$~eV); results computed before this choice were re-referenced ($\mu-\EF$ shifted by $-5.9$~meV, values at $\EF$ changed by $\le3\%$). The three compounds were Wannierized with the same projections and protocol (50 Bloch bands, outer and frozen windows $[-8,3]$ and $[-8,2]$~eV relative to $\EF$, $60\times60$ mesh); the DFT Fermi levels of NbS$_2$ and NbTe$_2$ are $-2.9935$ and $-1.7377$~eV (13-electron fillings $-2.9979$ and $-1.7449$~eV). Figure~\ref{S:fig:bands} compares the Wannier bands (computed with WannierTools~\cite{S:Wu2018}) with the fully relativistic DFT bands: between $-7$ and $+1.5$~eV they agree within 0.8~meV (median 0.05~meV), and within 0.5~meV for $|E-\EF|<0.5$~eV. The M saddle of the Nb pair lies at $-0.093$, $-0.207$ and $-0.176$~eV relative to the DFT Fermi levels of NbS$_2$, \NbSe{} and NbTe$_2$.

\subsection*{S1.3. Symmetrization}
Maximally localized Wannier models are symmetric only approximately: for \NbSe{} the raw model breaks $C_3$ by up to ${\sim}10^{-2}$~eV element-wise and by $4\times10^{-3}$~\AA{} in the position matrix $\bm r(\bm R)$, and the Wannier centres deviate from the ideal sites by $2\times10^{-4}$~\AA. These deviations barely affect the bands, but light-induced Berry-curvature hot spots amplify them into a spurious BCD of ${\sim}10^{-2}$~\AA{} under circular light, for which $C_3$ requires $\bm D=0$. Before any dressing we therefore (i) project the atomic positions onto $C_{3v}$-invariant ones, (ii) average $H(\bm R)$ and $\bm r(\bm R)$ over $C_{3v}$, and (iii) prune hoppings with symmetry-invariant site-pair norms. This reduces the spurious BCD under circular light to $\le5\times10^{-7}$~\AA{} ($\Heff$ at 10~eV, WannierBerri; Sec.~S4) and changes the bands by $\le0.15$~meV (NbSe$_2$) and $\le4\times10^{-4}$~eV (NbS$_2$, NbTe$_2$). The undriven symmetrized model gives $|\bm D|\le7\times10^{-12}$~\AA{} and $|\sigma_{xy}|\le10^{-14}\eh$ (WannierBerri, $4800^2$, 50~K); the FFEH evaluation of Sec.~S6 has a higher floor, $|\bm D|\le5.8\times10^{-5}$~\AA{} already at $\alpha=0$ ($2400^2$, maximum near the M saddle), which sets the floor of all FFEH symmetry checks. The DFT structure has the twelve operations of $D_{3h}$: the vertical mirror planes $\sigma_v^{(i)}$ contain the Nb--Se bonds (traces at $30^\circ$, $90^\circ$ and $150^\circ$ from $x$) and contain the twofold axes $C_2^{\prime(i)}$; $\sigma_v^{(1)}$ is $x\to-x$.

\subsection*{S1.4. Proximity exchange}
The exchange of a magnetic substrate is modelled, as in Ref.~\cite{S:Wicaksono2026}, by $H_{\rm ex}=\Delta\,\hat{\bm m}\cdot\bm s$ acting on the Se-$p$ Wannier pseudospin pairs of the bottom Se layer (orbitals 17--22), with $\bm s$ the Pauli matrices (eigenvalues $\pm1$, so a pure-spin orbital is split by $2\Delta$). It is added after symmetrization; being on site, it enters only the static Fourier component $H^{(0)}$. The main text uses $\hat{\bm m}=\hat{\bm z}$ and $\Delta>0$; $\Delta<0$ reverses $\hat{\bm m}$. With $\Delta=30$~meV and $\hat{\bm m}=\hat{\bm z}$, this convention reproduces the exchange-split bands of Ref.~\cite{S:Wicaksono2026} to within 0.12~meV at $\Gamma$, M and K for all three compounds. Chalcogen-$p$ states, shared by the two layers, shift by about $\Delta/2$ (at most $0.7\Delta$); the Nb pair is split by $\pm(2.3$--$2.8)$~meV at K and by about $\pm5$~meV at M in NbS$_2$ and \NbSe, but by $\pm13$~meV at M in NbTe$_2$ [Fig.~\ref{S:fig:bands}, lower row].

% =====================================================================
\section*{S2. Light--matter coupling, Floquet blocks and response functions}
% =====================================================================
\subsection*{S2.1. Coupling}
For normally incident light with the vector potential of Eq.~(1), $\bm e_1=(\cos\phi,\sin\phi,0)$ and $\bm e_2=\hat{\bm z}\times\bm e_1$, the cycle-averaged $|\bm A|^2=A_0^2$ is independent of $\chi$, so polarization scans are at fixed intensity. With $\alpha=eA_0/\hbar$, $E_0=\hbar\omega\alpha/e$ is the root-mean-square field [$E_0$(MV/cm)$=100\,\alpha$(\AA$^{-1}$)$\,\hbar\omega$(eV)]; the peak field of linear light is $\sqrt2E_0$. For electrons of charge $-e$ the driven Wannier Hamiltonian is~\cite{S:Ventura2017}
\begin{equation}
\begin{split}
H_{mn}(\bm R,t)={}&\left[H_{mn}(\bm R)+e\bm E(t)\cdot\bm{\mathcal D}_{mn}(\bm R)\right]\\
&\times e^{i(e/\hbar)\bm A(t)\cdot\bm d_{mn}(\bm R)},
\end{split}
\label{S:eq:coupling}
\end{equation}
with the Peierls vector $\bm d_{mn}(\bm R)=\bm R+\bm\tau_n-\bm\tau_m$ and the intracell dipole $\bm{\mathcal D}_{mn}(\bm R)=\bm r_{mn}(\bm R)-\bm\tau_m\delta_{mn}\delta_{\bm R0}$ (Hermitized), i.e., the position matrix minus the centres already contained in the Peierls phase. The dipole term restores interband couplings, such as Nb $d$--Se $p$, that Peierls phases alone miss. As a gauge test, results obtained with atomic centres and with all centres at the origin must coincide when the dipole term is included: they agree to $10^{-7}$--$10^{-5}$~eV, whereas Peierls phases alone differ by 3--4~meV (Table~\ref{S:tab:checks}).

\subsection*{S2.2. Fourier components}
Writing $(e/\hbar)\bm A(t)\cdot\bm d=z\cos(\omega t-\psi)$, with $z=\sqrt2\alpha[(\cos\chi\,\bm e_1\cdot\bm d)^2+(\sin\chi\,\bm e_2\cdot\bm d)^2]^{1/2}$ and $\psi={\rm atan2}(\sin\chi\,\bm e_2\cdot\bm d,\cos\chi\,\bm e_1\cdot\bm d)$, the Jacobi--Anger expansion gives $e^{iz\cos(\omega t-\psi)}=\sum_l c_le^{il\omega t}$ with $c_l=i^lJ_l(z)e^{-il\psi}$. With $H(t)=\sum_lH^{(l)}e^{il\omega t}$,
\begin{equation}
\begin{gathered}
H^{(l)}=H\,c_l+\hbar\omega\alpha\left(u\,c_{l-1}+v\,c_{l+1}\right),\\
u=\frac{p}{2i}-\frac q2,\quad v=-\frac{p}{2i}-\frac q2,\\
p=\sqrt2\cos\chi\,\bm e_1\cdot\bm{\mathcal D},\quad q=\sqrt2\sin\chi\,\bm e_2\cdot\bm{\mathcal D},
\end{gathered}
\label{S:eq:blocks}
\end{equation}
element by element in $(m,n,\bm R)$. The Floquet Hamiltonian is $[H_F(\bm k)]_{pq}=H^{(p-q)}(\bm k)+p\hbar\omega\,\delta_{pq}$, $p,q=-N_F,\dots,N_F$. The blocks agree with direct time quadrature to $10^{-16}$~eV and $H_F$ is Hermitian to $10^{-14}$~eV. For linear light ($\chi=0$; $\psi=0$ or $\pi$), $c_{-l}=c_l$, so the dipole contributions to $H^{(0)}$ cancel and $H^{(0)}$ depends on $\alpha$ only (checked to $2\times10^{-16}$~eV); the dipole parts of $H^{(\pm l)}$, by contrast, are opposite (Sec.~S4).

\subsection*{S2.3. Response functions}
The sheet anomalous Hall conductivity and the two-dimensional BCD are~\cite{S:Xiao2010,S:Sodemann2015,S:Wicaksono2026}
\begin{equation}
\begin{aligned}
\sigma_{xy}&=-\frac{e^2}{\hbar}\sum_n\int\!\frac{d^2k}{(2\pi)^2}f_n\Omega_n,\\
D_a\equiv D^{\rm 2D}_{az}&=\sum_n\int\!\frac{d^2k}{(2\pi)^2}f_n\,\partial_{k_a}\Omega_n\\
&=-\sum_n\int\!\frac{d^2k}{(2\pi)^2}(\partial_{k_a}f_n)\,\Omega_n ,
\end{aligned}
\label{S:eq:resp}
\end{equation}
in units of $e^2/h$ and \AA{} ($D^{\rm 2D}=c\,D^{\rm 3D}$). On a grid of $N_k$ points,
\begin{equation}
\begin{aligned}
\frac{\sigma_{xy}}{e^2/h}&=-\frac{2\pi}{A N_k}\sum_{\bm k,n}f_n\Omega_n,\\
D_a&=-\frac{1}{A N_k}\sum_{\bm k,n}(\partial_{k_a}f_n)\,\Omega_n,
\end{aligned}
\end{equation}
with $A$ the cell area. Berry curvatures are evaluated by Wannier interpolation including the full position matrix~\cite{S:Wang2006}; for the Floquet states,
\begin{gather}
\Omega_n=\bar\Omega_{nn}-[\mathcal Q_x,\bar A_y]_{nn}+[\mathcal Q_y,\bar A_x]_{nn}-i[\mathcal Q_x,\mathcal Q_y]_{nn},\nonumber\\
(\mathcal Q_a)_{nm}=\frac{(U^\dagger\partial_aH\,U)_{nm}}{\varepsilon_m-\varepsilon_n},
\label{S:eq:wysv}
\end{gather}
for $|\varepsilon_m-\varepsilon_n|\ge10^{-5}$~eV, with $(\mathcal Q_a)_{nm}=0$ otherwise (including $n=m$), so that for numerically degenerate bands only the group sum is gauge invariant (group treatment in Sec.~S6); where $\bar A=U^\dagger A^WU$ and $\bar\Omega=U^\dagger\Omega^WU$ are the Wannier-gauge connection and curvature rotated to the eigenbasis $U$ (photon sectors are diagonal in position). Occupations are Fermi--Dirac functions at temperature $T$ (Sec.~S5). A weak probe field $\bm E(t)={\rm Re}[\bm{\mathcal E}e^{i\omega_{\rm pr}t}]$ with $\omega_{\rm pr}\ll1/\tau$ then drives the nonlinear Hall current $\bm j(t)={\rm Re}[\bm j^{(0)}+\bm j^{(2)}e^{2i\omega_{\rm pr}t}]$, with second-harmonic amplitude $j^{(2)}_a=-\epsilon_{adc}\,e^3\tau D_{bd}\,\mathcal E_b\mathcal E_c/[2\hbar^2(1+i\omega_{\rm pr}\tau)]$ and a rectified part $j^{(0)}_a$ given by the same expression with $\mathcal E_c\to\mathcal E_c^*$~\cite{S:Sodemann2015}; this sets the device estimates of Sec.~S12.

\begin{table*}[t]
\caption{Consistency checks and benchmarks.}
\label{S:tab:checks}
\begin{ruledtabular}
\begin{tabular}{p{0.44\textwidth}p{0.44\textwidth}c}
Check & Result & Section \\ \hline
Centre convention (gauge), Peierls + dipole & $10^{-7}$--$10^{-5}$~eV (Peierls only: 3--4~meV) & S2 \\
Floquet blocks vs time quadrature; Hermiticity of $H_F$ & $10^{-16}$~eV; $10^{-14}$~eV & S2 \\
Undriven baseline (symmetrized), WannierBerri $4800^2$ & $|\bm D|\le7\times10^{-12}$~\AA, $|\sigma_{xy}|\le10^{-14}\eh$ & S1 \\
Undriven baseline, FFEH evaluation ($\alpha=0$, $2400^2$) & $|\bm D|\le5.8\times10^{-5}$~\AA{} (FFEH floor) & S1 \\
$k$ grid, WannierBerri, linear light, $\alpha=0.02$, 10~eV & $4800^2$ vs $8000^2$: 1.2\% rms, peak 0.2\% & S4 \\
Full Floquet, $A_0=0$ vs static; 40~eV vs $\Heff$ & $7\times10^{-15}$; $\sigma_{xy}$ within 1.3\% & S5 \\
Micromotion term, Eq.~\eqref{S:eq:mm}, vs exact Floquet states at 0.2~eV & error $\propto\alpha^4$ (median exponent 4.00); $\propto\alpha^2$ without it & S6 \\
FFEH vs full Floquet, 10 and 20~eV, pointwise $\Delta\Omega$ & within 0.1\% (without micromotion 7.3\%, 3.3\%) & S6 \\
FFEH vs full Floquet, 10~eV, integrated valve light part & ratio 1.000 (IQR 1.000--1.001) & S6 \\
FFEH vs full Floquet, 0.2~eV, 40 random $k$ & quasienergies $(1.5$--$2.3)\times10^{-7}$~eV (median); $\Delta\Omega$ 2--4\% (linear), 4--10\% (circ.+exch.) & S6 \\
FFEH, $\sigma_{xy}$ grid & $600^2$ vs $2400^2$: $<1\%$ rms & S6 \\
FFEH, $\bm D$ grid (0.2~eV) & $2400^2$ vs $4800^2$: $<1\%$ rms, peak 0.3\% & S6 \\
$C_3$-forbidden $\bm D$ (circular light), FFEH & $5.8\times10^{-5}$~\AA{} (at the FFEH floor) & S7 \\
Commutator correction to $\Heff$, linear light, dipole term on/off & $1.3$~meV (median norm, $\alpha=0.02\angi$) / $\le4\times10^{-18}$~eV & S4 \\
Horizontal mirror ($\Delta=0$): $\|[H,M_z]\|$; for comparison $\|[H,s_z]\|$ & $\le2\times10^{-9}$~eV; up to 0.42~eV & S5 \\
Dominant interband absorption onset ($\Delta=0$, full velocity, $T=0$) & 0.60~eV at $\EF$; 0.42~eV at $\mu-\EF=-0.205$~eV & S5 \\
\end{tabular}
\end{ruledtabular}
\end{table*}

% =====================================================================
\section*{S3. Polarization selection rules}
% =====================================================================
The Jones vector of Eq.~(1) is $\bm\varepsilon=\cos\chi\,\bm e_1-i\sin\chi\,\bm e_2$ [$\bm A(t)=\sqrt2A_0\,{\rm Re}(\bm\varepsilon e^{i\omega t})$], with Stokes parameters $S_1+iS_2=\cos2\chi\,e^{2i\phi}$ and $S_3=\sin2\chi$; $S_3>0$ ($\eta=+1$) for a field rotating counterclockwise seen from $+\hat{\bm z}$, i.e., photon angular momentum along $+\hat{\bm z}$. (For light incident from $+z$ this is negative helicity in the optics convention; we label states by the rotation sense in the sample frame.)

\paragraph*{Transformation of the drive.} Time reversal maps $\bm A(t)\to-\bm A(-t)$, which after a half-period shift is Eq.~(1) with $\chi\to-\chi$: $S_3$ is $\T$-odd and $S_{1,2}$ are $\T$-even. A rotation by $2\pi/3$ maps $\phi\to\phi+2\pi/3$, so $S_1\pm iS_2$ acquires $e^{\pm4\pi i/3}$ and $S_3$ is invariant. The mirror $\sigma_v^{(1)}$ ($x\to-x$) maps $\phi\to-\phi$ (mod $\pi$) and $\chi\to-\chi$.

\paragraph*{Transformation of the responses.} $\sigma_{xy}$ is a $\T$-odd rotation scalar. The BCD of Eq.~\eqref{S:eq:resp} is $\T$-even ($\Omega_n(-\bm k)=-\Omega_n(\bm k)$ and $\partial_{\bm k}$ changes sign) and transforms as an in-plane vector under rotations, $D_\pm\equiv D_x\pm iD_y\to e^{\pm2\pi i/3}D_\pm$. Under $\sigma_v^{(1)}$, $\Omega_z\to-\Omega_z$ and $k_x\to-k_x$, so $(D_x,D_y)\to(D_x,-D_y)$: under vertical mirrors $\bm D$ behaves as an axial vector.

\paragraph*{Selection rules.} To second order in the field, the responses are linear in the Stokes parameters. $\T$ and $C_3$ allow only $\sigma_{xy}\propto S_3$. For $D_+$, $C_3$ requires a combination carrying angular momentum $\equiv1\pmod3$ under rotations; among $S_0$, $S_3$ (0) and $S_1\pm iS_2$ ($\pm2$), only $S_1-iS_2$ qualifies, and it is $\T$-even. Hence $D_+=\beta(S_1-iS_2)=\beta\cos2\chi\,e^{-2i\phi}$ [Eq.~(4)]. Since $\sigma_v^{(1)}$ maps $D_+(\phi)\to D_+(-\phi)^*$, $\beta$ is real; for $\phi=0$ it forbids $D_y$. At higher orders, products such as $(S_1+iS_2)^2$ ($+4\equiv1$) give the $e^{4i\phi}$ harmonic; $\T$-odd products such as $S_3(S_1-iS_2)$ cannot enter $\bm D$, so $\bm D(\chi)=\bm D(-\chi)$ holds to all orders.

\paragraph*{With exchange.} An out-of-plane exchange ($\hat{\bm m}=\hat{\bm z}$) breaks $\sigma_h$ and $\T$ but keeps $C_3$; $\T$ maps $(\Delta,\eta)\to(-\Delta,-\eta)$, so $\sigma_{xy}(-\Delta,-\eta)=-\sigma_{xy}(\Delta,\eta)$ and $\bm D(-\Delta,-\eta)=\bm D(\Delta,\eta)$, both verified to $10^{-7}$. Circular light with $\hat{\bm m}\parallel\hat{\bm z}$ keeps $C_3$ and therefore still has $\bm D=0$. Because $\bm D$ is $\T$-even, terms odd in exactly one of the two $\T$-odd quantities $\Delta$ and $S_3$ are forbidden, such as $\Delta_z\alpha^2$ for linear light or $S_3(S_1-iS_2)$ without exchange. Products of both, such as $\Delta S_3(S_1-iS_2)\propto\Delta\sin4\chi$, are $\T$-even and not excluded by symmetry; they vanish for linear and circular light and were not computed for elliptical light with exchange. In-plane exchange breaks $C_3$ by itself.

% =====================================================================
\section*{S4. High-frequency limit}
% =====================================================================
\subsection*{S4.1. Effective Hamiltonian and WannierBerri}
To leading order in $1/\hbar\omega$,
\begin{equation}
\Heff=H^{(0)}+\sum_{\ell=1,2}\frac{[H^{(\ell)},H^{(-\ell)}]}{\ell\hbar\omega},
\label{S:eq:heffhf}
\end{equation}
with the sign verified against the full Floquet spectrum. The commutators are evaluated in real space (convolution over $\bm R$), so $\Heff(\bm R)$ is a tight-binding model that we write in Wannier90 format (original position matrix kept) and integrate with WannierBerri~\cite{S:Tsirkin2021} using the settings of Ref.~\cite{S:Wicaksono2026}: no imposed symmetry, $N_{\rm FFT}=80$, $N_{\rm div}=60$ ($4800^2$), Fermi--Dirac smoothing at 50~K. Figure~\ref{S:fig:hf}(a) shows the convergence: $4800^2$ reproduces $8000^2$ within 1.2\% (rms) and the peak within 0.2\%. For linear light, $c_{-l}=c_l$, so the Peierls parts of opposite harmonics coincide, $P_{-l}=P_l$, while the dipole parts are opposite, $B_{-l}=-B_l$ [Eq.~\eqref{S:eq:blocks} with $H^{(l)}=P_l+B_l$]; the commutator therefore reduces to the mixed term $[H^{(l)},H^{(-l)}]=-2[P_l,B_l]$, which does not vanish. Peierls phases alone give a vanishing commutator ($\le4\times10^{-18}$~eV at random $\bm k$). With the dipole term, at $\alpha=0.02\angi$ the commutator term of Eq.~\eqref{S:eq:heffhf} has a median norm of 1.3~meV over 30 random $\bm k$ (maximum 2.3~meV), against 4.2~meV for the light-induced change of $H^{(0)}$; it lowers the saddle at M$_1$ ($k_y$ axis) by 1.8~meV and those at M$_2$, M$_3$ by 1.4~meV, reducing their light-induced splitting from 2.5 to 2.1~meV. Because $B_l\propto\hbar\omega\alpha$, this term, like $H^{(0)}$, is independent of $\hbar\omega$ at fixed $\alpha$ (to $2\times10^{-18}$~eV between 6 and 10~eV), so the linear-light $\Heff$ is independent of $\hbar\omega$: runs at 6, 10 and 15~eV agree to $10^{-16}$. The commutators are evaluated numerically for every polarization with the dipole term included, so this contribution is contained in all high-frequency results.

\subsection*{S4.2. Light only}
\begin{figure*}[t]
\includegraphics[width=\textwidth]{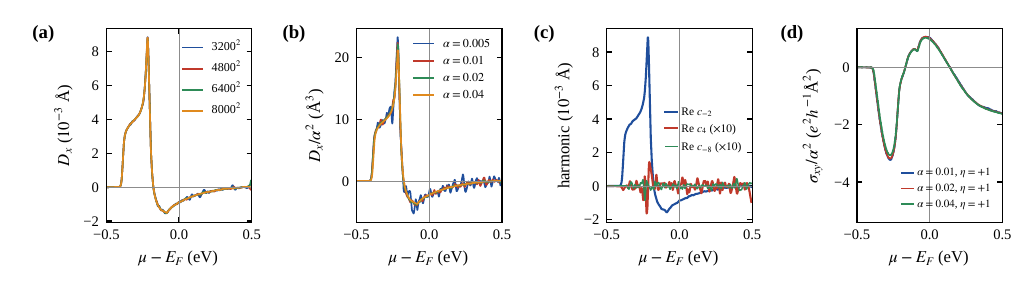}
\caption{High-frequency limit, light only ($\Heff$ at 10~eV, WannierBerri, 50~K). (a) $k$-grid convergence of $D_x$ for linear light, $\alpha=0.02\angi$ ($N_{\rm FFT}=80$, $N_{\rm div}=40$--100; Fermi--Dirac smoothed). (b) $D_x/\alpha^2$ for $\alpha=0.005$--$0.04\angi$. (c) Harmonics of $D_+(\phi)=\sum_nc_ne^{in\phi}$ fitted from $\phi=0$--$30^\circ$ ($\alpha=0.02\angi$; $c_4$, $c_{-8}$ magnified 10 times). (d) $\sigma_{xy}/\alpha^2$ for circular light, $\eta=+1$ (solid) and $-\sigma_{xy}(\eta=-1)/\alpha^2$ (dotted, indistinguishable).}
\label{S:fig:hf}
\end{figure*}

Linear light ($\alpha=0.02\angi$) gives a peak $D_x=8.8\times10^{-3}$~\AA{} at $\mu-\EF=-0.216$~eV and $D/\alpha^2\approx22$~\AA$^3$, with a quartic correction of $-4\%$ at $\alpha=0.04\angi$ [Fig.~\ref{S:fig:hf}(b)]; $|\sigma_{xy}|\le7\times10^{-9}\eh$ and the mirror-forbidden $|D_y|\le8\times10^{-7}$~\AA. The angle dependence follows $D_+=\beta e^{-2i\phi}$ with higher harmonics $\le2\%$ of $\beta$, at the noise floor [Fig.~\ref{S:fig:hf}(c)]. Circular light gives $\sigma_{xy}/\alpha^2\approx-3.2\ehA$ at $\mu-\EF\approx-0.28$~eV with $\sigma(+\eta)=-\sigma(-\eta)$ to $10^{-8}$ [Fig.~\ref{S:fig:hf}(d)] and $|\bm D|\le5\times10^{-7}$~\AA. Elliptical light obeys $\sigma=\sin2\chi\,\sigma_{\rm circ}$ and $\bm D=\cos2\chi\,\bm D_{\rm lin}$ within 1\%, with $\bm D(\chi)=\bm D(-\chi)$ to numerical precision. Full Floquet at 10~eV gives a BCD larger by a factor 1.07 ($0.94\times10^{-2}$~\AA; Sec.~S5).

\subsection*{S4.3. Light combined with exchange}
\begin{figure*}[t]
\includegraphics[width=\textwidth]{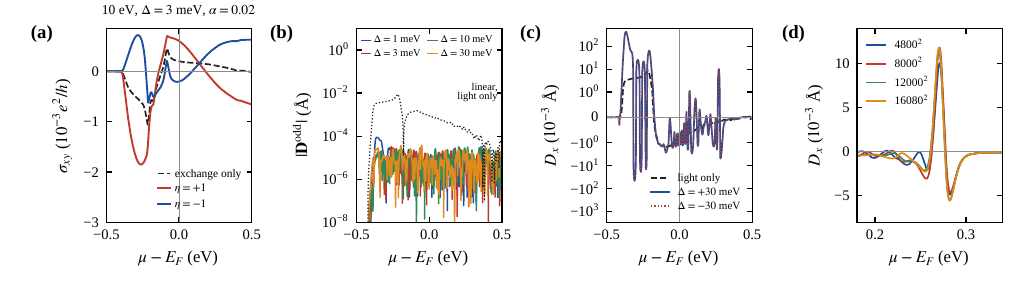}
\caption{High-frequency limit with exchange ($\Heff$ at 10~eV, $\alpha=0.02\angi$, $4800^2$, 50~K). (a) Configuration B: $\sigma_{xy}$ for $\Delta=3$~meV without light and with $\eta=\pm1$. (b) Configuration A: magnitude of the helicity-odd BCD, $|\bm D(\eta=+1)-\bm D(\eta=-1)|/2$, for in-plane exchange $\Delta=1$--30~meV, compared with $|D_x|$ for linear light alone. (c) Configuration C: $D_x$ for $\Delta=\pm30$~meV (symmetric-log scale) and for light alone. (d) Grid convergence of the resonance at $\mu-\EF=+0.27$~eV.}
\label{S:fig:hfex}
\end{figure*}

Three configurations were studied with exchange on the bottom Se layer ($\Delta=1$--30~meV, $\alpha=0.02\angi$, $4800^2$; Fig.~\ref{S:fig:hfex}).

\emph{(B) $\hat{\bm m}\parallel\hat{\bm z}$ + circular light: Hall valve.} The response is additive, $\sigma_{xy}\approx\Delta\,s(\mu)+\eta\alpha^2\ell(\mu)$, with the light part equal to the light-only value and cross terms $\le4\times10^{-5}\eh$ [Fig.~\ref{S:fig:hfex}(a)]. The exchange response, $\max_\mu|\sex|/\Delta=(3.57,\allowbreak\,3.54,\allowbreak\,3.47,\allowbreak\,3.23)\times10^{-4}\eh$ per meV for $\Delta=1$, 3, 10 and 30~meV, is peak based and not evaluated at $\EF$. A peak-based characteristic exchange scale, $\Delta_{\rm pk}=\max_\mu|\slight|/(\max_\mu|\sex|/\Delta)$, compares maxima at different $\mu$ and is not a sign-reversal threshold; it is 0.9, 3.6 and 13.9~meV for $\alpha=0.01$, 0.02 and $0.04\angi$ ($\Heff$) and $\approx2.4$~meV at $\alpha=0.02\angi$ with full Floquet. The $\EF$-based crossover of the main text, $\sex(\EF;\Delta^*)=|\slight(\EF)|$, gives $\Delta^*\approx6$~meV at 10~eV and $\alpha=0.02\angi$.

\emph{(A) in-plane $\hat{\bm m}$ + circular light.} Circular light does not act as an optical $m_z$ in a product law with in-plane exchange: the helicity-odd BCD [Fig.~\ref{S:fig:hfex}(b)] stays below $9\times10^{-5}$~\AA{} for all $\Delta=1$--30~meV, uncorrelated with $\Delta$ (numerical floor), compared with ${\sim}10^{-2}$~\AA{} for linear light.

\emph{(C) $\hat{\bm m}\parallel\hat{\bm z}$ + linear light.} There is no broad enhancement: away from hot spots $D$ equals its light-only value, and it is even in $\hat{\bm m}$ [Fig.~\ref{S:fig:hfex}(c)]. One converged resonance appears at $\mu-\EF=+0.270$~eV, where $D=+1.18\times10^{-2}$/$-5.5\times10^{-3}$~\AA{} over ${\sim}10$~meV (50 times the light-only value), stable from $4800^2$ to $16080^2$ [Fig.~\ref{S:fig:hfex}(d)]. A hot spot at $-0.37$~eV remains unresolved up to $16080^2$ (alternating sign, $|D|\lesssim0.15$~\AA{} and shrinking with the grid). These configurations are not used in the main text beyond the statement that they give no comparable effect.

% =====================================================================
\section*{S5. Full Floquet steady state and occupations}
% =====================================================================
\subsection*{S5.1. Method}
The full Floquet Hamiltonian (Sec.~S2, $N_F=2$ unless stated) is diagonalized at each $\bm k$; a Floquet state $a$ has quasienergy $\varepsilon_a$ and sector weights $w_a^p$. For a weak, wide-band fermionic bath at chemical potential $\mu$~\cite{S:Seetharam2015,S:Dehghani2014}, $n_a=\sum_pw_a^pf(\varepsilon_a-p\hbar\omega-\mu)$; for a bath of half-bandwidth $W$, golden-rule rate balance gives $n_a=\sum_pw_a^p\rho_pf_p/\sum_pw_a^p\rho_p$ with $\rho_p=\Theta(W-|\varepsilon_a-p\hbar\omega-\mu|)$. The dressed occupation is $n_a=f(\varepsilon_a-\mu)$. An energy-filtered bath reproduces it when only the central channel $p=0$ contributes and its coupling is nonzero; $2W<\hbar\omega$ ensures at most one channel for a given quasienergy but does not by itself select $p=0$, which depends on the bath alignment and the quasienergy convention. Each Floquet family $\{\varepsilon_a+m\hbar\omega\}$ is counted once: we keep the copy whose mean sector index lies in $[-\tfrac12,\tfrac12)$ or, at low frequency, the copy that maximizes the total overlap with the sector-0 bands of $H^{(0)}$ (Hungarian assignment), and $\varepsilon_a$ is the quasienergy of that copy. Berry curvatures follow Eq.~\eqref{S:eq:wysv} in extended space. Benchmarks: at $A_0=0$ the result equals the static one to $7\times10^{-15}$; at 40~eV $\sigma_{xy}$ agrees with $\Heff$ within 1.3\%; a $2400^2$ grid takes 7.6~min on 128 cores.

\subsection*{S5.2. Results and limits}
At 10~eV with dressed occupations, full Floquet gives a linear-light BCD 1.07 times the $\Heff$ value, and its circular-light (helicity-odd) part of the Hall-valve $\sigma_{xy}$ is 0.67 times the $\Heff$ value (interquartile range of the pointwise ratio 0.55--0.77) [Fig.~\ref{S:fig:ness}(b)]; the FFEH reproduces full Floquet within the reported numerical accuracy (Sec.~S6). A wide-band bath produces photon-assisted filling of replicas: at 10~eV it shifts the linear-light $D_x$ by a nearly constant offset that persists below the band bottom and changes its extremum from $+9.4\times10^{-3}$ to $-7.5\times10^{-3}$~\AA{} [Fig.~\ref{S:fig:ness}(c)]. For $\hbar\omega$ below the total bandwidth the integrated steady state does not converge: at 3~eV (resonant) $\sigma_{xy}$ is 14, 7.2 and 2.7$\eh$ for $600^2$, $1200^2$ and $2400^2$ [Fig.~\ref{S:fig:ness}(a)], and at 0.3~eV it depends on $N_F$ and the broadening. The cause is the dense folding of a ${\sim}10$~eV-wide manifold. We estimate that a full Floquet--Keldysh (Bastin) treatment would cost ${\sim}10^4$ times more in our implementation; it was not attempted.

\subsection*{S5.3. Occupation model used in the main text}
\begin{figure*}[t]
\includegraphics[width=\textwidth]{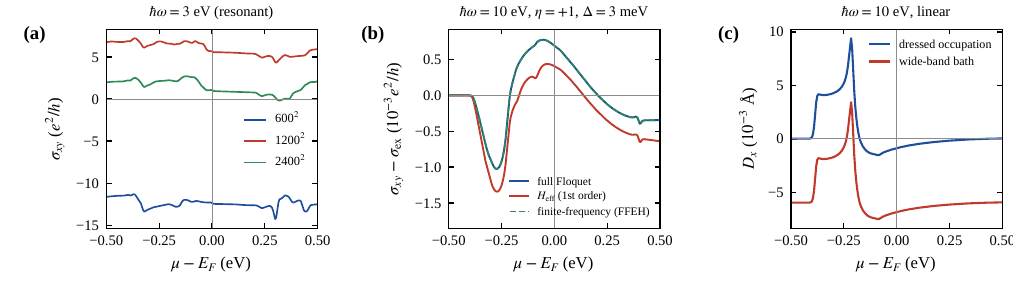}
\caption{Full Floquet steady state ($\alpha=0.02\angi$). (a) Non-convergence at the resonant $\hbar\omega=3$~eV (circular light, $\Delta=3$~meV, wide-band occupations): $\sigma_{xy}$ for three grids. (b) 10~eV, $\eta=+1$, $\Delta=3$~meV: light-induced part $\sigma_{xy}-\sex$ from full Floquet (dressed), the first-order $\Heff$ and the FFEH (dashed, on top of full Floquet). (c) 10~eV, linear light: $D_x$ with dressed occupations and with a wide-band bath.}
\label{S:fig:ness}
\end{figure*}

The main text assumes Fermi--Dirac occupations of the dressed bands at the electron temperature. Off resonance it rests on the following considerations, which support but do not establish it: (i) without exchange the model keeps the horizontal mirror $\sigma_h$ [$\|[H(\bm k),M_z]\|\le2\times10^{-9}$~eV, with $M_z$ the spinful mirror acting on orbitals, layers and spin], although $s_z$ itself is not conserved ($\|[H(\bm k),s_z]\|$ up to 0.42~eV). The two Ising-split Nb branches lie in opposite mirror sectors ($M_z=\pm i$), and the in-plane position and velocity operators are mirror even, so in-plane electric-dipole transitions between the branches are forbidden. The dominant interband absorption, from Se~$p$ to Nb, sets in at 0.60~eV at $\EF$ and at 0.42~eV at the BCD peak ($\mu-\EF=-0.205$~eV), above the drive frequencies. Layer-selective exchange breaks $\sigma_h$ and weakly activates the low-energy transitions: the interband weight of ${\rm Re}\,\sigma_{xx}$ below 0.3~eV, relative to that between 0.6 and 0.9~eV, is $3\times10^{-6}$ ($7\times10^{-7}$) for $\Delta=3$~meV and $5\times10^{-4}$ ($2\times10^{-4}$) for 30~meV at $\EF$ (at the BCD peak), against $10^{-18}$ without exchange, and the onsets move down by at most 0.03~eV. These numbers use the full Wannier-interpolated velocity $\hbar\bm v=\partial_{\bm k}H^W-i[\bm A^W,H^W]$, with $\bm A^W$ the Wannier centres plus the residual position matrix of Eq.~\eqref{S:eq:coupling} (its term changes the weights between 0.6 and 1~eV by less than 10\% and leaves the onsets unchanged), zero-temperature occupations at $\mu$, a $96^2$ grid and 0.01~eV bins; the onset is where the interband spectrum first exceeds 1\% of its mean between 0.7 and 0.8~eV. At finite temperature the occupations smear these onsets by a few $k_BT$; (ii) the Nb--Se-$p$ near-resonance at $\gtrsim0.35$~eV connects two occupied bands (Nb at $-0.34$~eV, Se $p$ at $-0.69$~eV relative to $\EF$) and is Pauli blocked, also at the BCD peak ($\mu-\EF\approx-0.2$~eV), where both states remain below $\mu$; (iii) the filling of photon replicas by a wide-band bath, which occurs even below the band bottom [Fig.~\ref{S:fig:ness}(c)], is a reservoir-dependent effect that we do not expect for a monolayer on an insulating substrate; and (iv) below the dominant onset we neglect the weak exchange-activated interband absorption and approximate the absorbed power by its intraband (Drude) part, treated through the electron temperature (Sec.~S12): at fixed $\mu$, the light-induced Hall conductivity at $\EF$ changes by 4\% up to 300~K and by 12\% up to 550~K, and the BCD peak (maximum over $\mu$) decreases by 41\% and 55\%, respectively (Sec.~S9). The occupations are assigned to the eigenstates of $\Heff$, which connect continuously to the undriven bands, so no zone folding of quasienergies enters. Dressed occupations remain an assumption; a kinetic calculation for a specific substrate and pulse would be needed to establish them.

\subsection*{S5.4. Sensitivity to the occupations}
Two alternatives to dressed Fermi--Dirac occupations at fixed $\mu$ indicate how strongly the results depend on the occupation model; both are evaluated at 50~K with the FFEH of the main text. Let $\varepsilon^0_n$ and $\Omega^0_n$ be the undriven bands and curvatures (same exchange), and $P(n)$ the undriven state of maximal overlap with the dressed state $n$; energy ordering gives the same pairing for all quantities below.

\textit{Fixed particle density.} At fixed $\mu$ the drive changes the electron number by $\delta n=\sum_n\int_{\bm k}[f(\varepsilon_n)-f(\varepsilon^0_n)]$; fixing the density instead shifts the chemical potential, to linear order, by $\delta\mu=-\delta n/N(\mu)$. At $\EF$, $\delta\mu=-0.04$ and $-0.16$~meV for circular light at $\alpha=0.01$ and $0.02\angi$, equal for both helicities within $3\times10^{-4}$. Being helicity even and of order $\alpha^2$, it changes $\slight(\EF)$ only at order $\alpha^4$, by $2\times10^{-4}$ and $8\times10^{-4}$ (relative). For linear light at $\alpha=0.02\angi$, $\delta\mu=-0.16$~meV at $\EF$, where $D_x$ changes by $2\times10^{-3}$, and $+1.3$~meV at the BCD peak. There $\partial_\mu D_x=0$, so a sample at the peak density changes only at second order, by $\delta D_x/D_x\simeq\delta\mu^2\,\partial^2_\mu D_x/(2D_x)\approx-0.3\%$ ($-0.24$ to $-0.28\%$ from three- and five-point estimates of the curvature on the 5~meV grid). In the equilibrium chemical potential that labels the density, the peak moves by $-1.3$~meV, i.e., it is reached with $4\times10^{-3}$ more holes per unit cell, against the 0.69 holes per Nb needed to reach it.

\textit{Inherited occupations.} If each dressed state keeps the occupation of the undriven state it connects to, as for adiabatic switching without relaxation, the Hall conductivity changes by $-\sigma^{\rm pop}$, with $\sigma^{\rm pop}=-(e^2/\hbar)\sum_n\int_{\bm k}[f(\varepsilon_n)-f(\varepsilon^0_{P(n)})]\,\Omega_n$. This population term is confined to the Fermi surface (we sum over $\bm k$ points with an undriven band within 65~meV of $\mu$) and agrees within 0.5\% with its linearized form $(e^2/\hbar)\sum_n\int_{\bm k}(-\partial_\varepsilon f)\,\delta\varepsilon_n\Omega^0_n$, in which the light-induced band shifts $\delta\varepsilon_n=\varepsilon_n-\varepsilon^0_{P(n)}$ act on the equilibrium curvature. Without exchange, time reversal makes the helicity-odd part of $\delta\varepsilon_n(\bm k)$ odd in $\bm k$, like $\Omega^0_n$, so the term contributes to $\slight$. Through order $\alpha^2$ it involves only the band shifts and the undriven curvature; the micromotion correction, which changes curvatures but not quasienergies, enters it only at order $\alpha^4$. Its helicity-odd part at $\EF$ is $-1.05\times10^{-4}$ and $-4.23\times10^{-4}\eh$ at $\alpha=0.01$ and $0.02\angi$ ($600^2$ and $1200^2$ grids; at $0.02\angi$ the two grids agree within 0.1\%), i.e., $-1.05\ehA$ times $\alpha^2$: 27\% of $c_2=-3.85\ehA$ and 22\% of $\slight(\EF)$ at $0.02\angi$. The remaining geometric part, $-2.80\ehA$, has the same sign. Inherited occupations therefore leave the sign of $\slight$ unchanged and give $\Delta^*=1.1$, 4.4 and 20~meV at $E_0=0.1$, 0.2 and 0.4\MVcm{} at leading order, against 1.5, 6.1 and 29~meV. The two prescriptions differ in whether the light-induced band shifts repopulate the Fermi surface during the pulse, which only a kinetic calculation can decide.

% =====================================================================
\section*{S6. Finite-frequency Floquet effective Hamiltonian with micromotion}
% =====================================================================
\subsection*{S6.1. Effective Hamiltonian}
In the eigenbasis of $H^{(0)}(\bm k)$ (energies $\varepsilon_n$), the states $|n,0\rangle$ of the central Floquet sector couple to $|l,p\rangle$ ($p\ne0$, energy $\varepsilon_l+p\hbar\omega$) through $H^{(-p)}_{nl}$. Decoupling the central sector to second order (Floquet--L\"owdin, or Schrieffer--Wolff~\cite{S:Bukov2016,S:Mikami2016,S:Eckardt2015}; for a related low-frequency construction see Ref.~\cite{S:Vogl2020}) gives Eq.~(2) of the main text,
\begin{equation}
\begin{aligned}
[\Heff]_{nm}&=\varepsilon_n\delta_{nm}+\tfrac12\sum_{p\ne0}\sum_lH^{(-p)}_{nl}H^{(p)}_{lm}\\
&\quad\times\left[g(\varepsilon_n-\varepsilon_l-p\hbar\omega)+g(\varepsilon_m-\varepsilon_l-p\hbar\omega)\right],\\
g(x)&=\frac{x}{x^2+\gamma^2},
\end{aligned}
\label{S:eq:ffeh}
\end{equation}
with the blocks of Eq.~\eqref{S:eq:blocks} ($N_F=2$). For $\hbar\omega\to\infty$, $g\to-1/(p\hbar\omega)$ and Eq.~\eqref{S:eq:ffeh} reduces to Eq.~\eqref{S:eq:heffhf}. At 10~eV its quasienergies match full Floquet to $7\times10^{-7}$~eV, and the light-induced Berry curvature without micromotion is within 6\% (first-order $\Heff$: 14\%); the remaining 6\% is the micromotion term below.

\subsection*{S6.2. Regularization}
The function $g$, the real part of $1/(x+i\gamma)$, regularizes small denominators with a broadening $\gamma=\hbar/\tau=26$~meV set by the scattering time; it is a regularization motivated by $\tau$, not a full self-energy, and its effect is quantified in Sec.~S6.5 and Table~\ref{S:tab:robust}. With $\gamma=1$~meV, resonant pairs of remote occupied bands produce second-order shifts of up to ${\sim}0.2$~eV that reorder bands; in a $60^2$ test at 0.3~eV this gave a $C_3$-forbidden $D_x=5.5\times10^{-3}$~\AA{} and a light part 200 times the 10~eV value. With $\gamma=26$~meV the shifts stay $\le29$~meV and the forbidden $D$ returns to the level that the same code gives at $\alpha=0$ (${\sim}5\times10^{-5}$~\AA; Sec.~S1). Near $\EF$ the expansion is safe at 0.2~eV (Sec.~S6.6), less safe at 0.3~eV ($\le0.6$ in 2.7\% of $\bm k$, for the occupied pair Nb at $-0.34$~eV and Se $p$ at $-0.69$~eV) and breaks down at 0.45~eV (exact resonances in 3.4\% of $\bm k$).

\subsection*{S6.3. Micromotion correction}
That the Berry curvature of Floquet states contains, besides that of the Floquet Hamiltonian, a contribution of the micromotion was shown in general in Ref.~\cite{S:Weinberg2017}; here we derive it to second order in the field, in the Wannier gauge. The derivation is for $\gamma\to0$; for finite $\gamma$ the generator below leaves a residual coupling $H^{(p)}_{ln}\gamma^2/(x^2+\gamma^2)$, small when the detuning $|x|\gg\gamma$, and we evaluate the same expressions with the regularized denominators (halving $\gamma$ changes $\slight(\EF)$ by 0.1\%).

\emph{Floquet state.} Write $H_F=H_F^{0}+V$, with $[H_F^{0}]_{pp}=H^{(0)}+p\hbar\omega$ and $V$ the blocks $H^{(p-q)}$, $p\ne q$. Because $c_l\propto J_l(z)$ in Eq.~\eqref{S:eq:blocks}, $H^{(p)}=O(\alpha^{|p|})$. With $H^{(0)}=\sum_n\varepsilon_n|n\rangle\langle n|=U_0{\rm diag}(\varepsilon)U_0^\dagger$ and $P$ the projector onto the central sector, the anti-Hermitian generator
\begin{equation}
\begin{aligned}
S&=\sum_{p\ne0}\left(T_p\otimes|p\rangle\langle0|-T_p^\dagger\otimes|0\rangle\langle p|\right),\\
T_p&=\sum_{l,n}|l\rangle\,H^{(p)}_{ln}\,g(\varepsilon_n-\varepsilon_l-p\hbar\omega)\,\langle n|
\end{aligned}
\label{S:eq:sw}
\end{equation}
removes the coupling of the central sector at first order, $[H_F^{0},S]=-(PVQ+QVP)$ for $\gamma\to0$ ($Q=1-P$), and $P\,e^{-S}H_Fe^{S}P=P(H_F^0+\tfrac12[V,S])P$ is Eq.~\eqref{S:eq:ffeh}. The next terms are $O(\alpha^4)$: a third-order process $0\to p\to q\to0$ carries $H^{(p)}H^{(q-p)}H^{(-q)}=O(\alpha^{|p|+|q-p|+|q|})$ with $p\ne q$, $p,q\ne0$. The central Floquet state of band $n$ is then
\begin{equation}
\begin{aligned}
|\Phi_n\rangle&\simeq e^{S}\big(|\psi_n\rangle\otimes|0\rangle\big)\\
&=\left(1-\tfrac12M\right)|\psi_n\rangle\otimes|0\rangle+\sum_{p\ne0}T_p|\psi_n\rangle\otimes|p\rangle+\dots,\\
M&=\sum_{p\ne0}T_p^\dagger T_p ,
\end{aligned}
\label{S:eq:phi}
\end{equation}
with $\psi_n$ the eigenvectors of $\Heff$. The omitted terms (higher-order generators) are $O(\alpha^3)$ in sectors $\pm1$, $O(\alpha^2)$ in sectors $|p|\ge2$ and $O(\alpha^4)$ in sector 0, and change the Berry connection only at $O(\alpha^4)$.

\emph{Berry connection.} In the Wannier gauge a Bloch state with coefficient vector $c(\bm k)$ has $A_a=i\,c^\dagger\partial_ac+c^\dagger A^W_ac$. For a Floquet state the time average sums this over the sectors, since position is diagonal in $p$; this is the connection whose curl Eq.~\eqref{S:eq:wysv} evaluates. With $u_p=T_p\psi_n$, sector $p\ne0$ contributes $\langle u_p|A^W_a|u_p\rangle$ and $i\langle u_p|\partial_au_p\rangle=\tfrac i2\partial_a\|u_p\|^2-{\rm Im}\langle u_p|(\partial_aT_p)\psi_n\rangle-{\rm Im}\langle u_p|T_p\partial_a\psi_n\rangle$. Sector 0 contributes $A_a[\psi_n]-\tfrac i2\partial_a\langle\psi_n|M|\psi_n\rangle+{\rm Im}\langle M\psi_n|\partial_a\psi_n\rangle-{\rm Re}\langle A^W_a\psi_n|M\psi_n\rangle+O(\alpha^4)$. The imaginary total derivatives cancel (normalization), and $\sum_p{\rm Im}\langle u_p|T_p\partial_a\psi_n\rangle={\rm Im}\langle M\psi_n|\partial_a\psi_n\rangle$ cancels the derivative of $\psi_n$. The result is $A_{n,a}=A_a[\psi_n]+\delta A_a+O(\alpha^4)$ with the micromotion term
\begin{equation}
\begin{split}
\delta A_a=\sum_{p\ne0}\Big[&-{\rm Im}\langle T_p\psi_n|(\partial_aT_p)\psi_n\rangle\\
&+\langle T_p\psi_n|A^W_a|T_p\psi_n\rangle\\
&-{\rm Re}\langle A^W_a\psi_n|T_p^\dagger T_p\psi_n\rangle\Big]:
\end{split}
\label{S:eq:mm}
\end{equation}
the connection carried by the $\bm k$ dependence of the sideband admixture, the intracell position of the admixed components, and the weight they remove from the central sector. The expression is invariant under these band-phase transformations, $\psi_n\to e^{i\theta(\bm k)}\psi_n$ and the phases of $|n\rangle$ ($T_p$ is an operator), so in the fixed Wannier gauge $\delta\bm A$ is a well-defined vector field, and the Floquet Berry curvature is $\Omega_n=\Omega_n[\Heff]+\partial_x\delta A_y-\partial_y\delta A_x+O(\alpha^4)$, with $\Omega_n[\Heff]$ from Eq.~\eqref{S:eq:wysv} and the curl by central differences ($\delta k=2\times10^{-5}$~\AA$^{-1}$). The derivative $\partial_aT_p$ is the central difference of $T_p$ as an operator in the orbital (Wannier-gauge) basis, which does not depend on the eigenvector phases. The broadened denominators $g$ of $\Heff$ are used in $T_p$ as well.

\emph{Degenerate bands.} At symmetry-protected degeneracies (at TRIMs when $\T$ holds, and on $\Gamma$--M lines invariant under a preserved vertical mirror, which anticommutes with $\sigma_h$) and at numerically unresolved near-degeneracies, individual-band connections are not uniquely defined; we therefore assign to each group of bands closer than 2~meV the curl of the group-summed connection, shared equally (a non-Abelian trace). The trace is exact for exactly degenerate groups; for groups split by up to 2~meV the equal sharing is an approximation, confined to narrow strips around the $\Gamma$--M lines. Repeating the production runs with thresholds of 0.5, 1 and 4~meV changes the BCD peak by $+0.9$, $+0.9$ and $-1.7\%$, $D_x(\EF)$ by at most 0.1\% and $\slight(\EF)$ by at most 0.01\%, turns the direction of $\bm D$ at $\phi=20^\circ$ by less than $0.2^\circ$, and keeps the $C_3$-forbidden $\bm D$ for circular light at the floor ($5.7\times10^{-5}$~\AA); the changes are confined to $-0.38<\mu-\EF<-0.19$~eV, around the M saddle (Table~\ref{S:tab:robust}). Without grouping, circular light produces a $C_3$-forbidden $D$ of $6\times10^{-3}$ and $1.4\times10^{-2}$~\AA{} at $\alpha=0.01$ and $0.02\angi$ ($2400^2$), and the angle and ellipticity laws of $\bm D$ are violated. The band velocities entering $\partial_af_n=f'(\varepsilon_n)\,v_{n,a}$ are $v_{n,a}=\partial_{k_a}\varepsilon_n=\langle\psi_n|\partial_{k_a}\Heff|\psi_n\rangle$ (the convention of Eq.~\eqref{S:eq:resp}, without $1/\hbar$), with $\partial_{k_a}\Heff$ from central differences. The cost is ${\approx}20$~ms per $\bm k$ point per core.

\subsection*{S6.4. Validation}
At 10 and 20~eV the pointwise light-induced Berry curvature agrees with full Floquet within 0.1\%, for linear light and for circular light with $\Delta=3$~meV (7.3\% and 3.3\% without the micromotion term); the integrated light part of the valve at 10~eV has the ratio 1.000 (IQR 1.000--1.001) to full Floquet [Fig.~\ref{S:fig:ness}(b)]. At 0.2~eV, where the integrated full-Floquet steady state fails, we compare state by state at 40 random $\bm k$ points for the Nb pair ($\alpha=0.01\angi$), using full Floquet with overlap-based state selection, converged for $N_F=4$--8 [Fig.~\ref{S:fig:ffeh}(a)]. Quasienergies agree to $(1.5$--$2.3)\times10^{-7}$~eV (median) and $\le4\times10^{-5}$~eV (maximum). The median $|\Delta\Omega|$ agrees within 1\%; pointwise, the median relative difference is 2--4\% (linear) and 4--10\% (circular with 3~meV exchange) when no other Floquet level lies within 1~meV. The outliers, ${\approx}7\%$ of band points, all have such a level, where unbroadened full Floquet is ill conditioned; reducing $\gamma$ in the FFEH to 5 or 1~meV worsens the agreement, so the residual discrepancy is not removed by reducing $\gamma$.

We checked Eqs.~\eqref{S:eq:phi} and \eqref{S:eq:mm} independently against exact Floquet states ($N_F=5$) at 40 random $\bm k$ points for the Nb pair at 0.2~eV with $\gamma\to0$, for $\alpha=0.0025$--$0.02\angi$, linear light ($\Delta=0$) and circular light of both helicities with $\Delta=3$~meV. The reference curvature does not use Eq.~\eqref{S:eq:wysv}: it is the plaquette Berry phase of the extended-space coefficient vector (side $10^{-4}\angi$) plus the curl of $\langle\Phi_n|A^W|\Phi_n\rangle$, and it reproduces the static curvature of the FFEH to $3\times10^{-6}$. The error of the light-induced curvature of the FFEH scales as $\alpha^4$ (median exponent 4.00, interquartile range 3.94--4.08) at 83--86\% of the band points; the others have another Floquet level within the plaquette, where the unbroadened reference is ill conditioned. Without $\delta\bm A$ the error scales as $\alpha^2$ and is 1.5 (linear) and 39 (circular, helicity-odd part) times the light-induced curvature [Fig.~\ref{S:fig:ffeh}(d)]; with the first term of Eq.~\eqref{S:eq:mm} only, it still scales as $\alpha^2$ (0.6--1.0 times the curvature for linear light), so the position-matrix terms are required. The code reproduces an independent implementation of Eq.~\eqref{S:eq:mm} to $10^{-9}$~\AA$^2$ (group sums on the 2~meV groups).

Two consequences matter for the mid-infrared results. First, the micromotion term is not a small correction at 0.2~eV. For the helicity-odd curvature it cancels that of $\Heff$ to within a few per cent (median pointwise ratio $-1.05$ for the Nb pair), and after integration ($\sigma_{\rm odd}$ at $\Delta=3$~meV, Sec.~S9) it reverses the sign of the light-induced Hall conductivity over about 80\% of the $\mu$ window: $\slight(\EF)=+2.8\times10^{-3}$ from $\Heff$ alone against $-4.1\times10^{-4}\eh$ with $\delta\bm A$ at $\alpha=0.01\angi$, and $+1.12\times10^{-2}$ against $-1.87\times10^{-3}\eh$ at $0.02\angi$ ($96^2$ grid). The sign change relative to 10~eV (Sec.~S9) is therefore a micromotion effect. Second, because the helicity-odd curvature is a small difference of two $O(\alpha^2)$ terms, the $O(\alpha^4)$ error is large relative to it: the ratio of the median absolute error to the median magnitude of the light-induced curvature, over the band points with $\alpha^4$ error scaling, is 0.36 at $\alpha=0.01\angi$ and 1.4 at $0.02\angi$ (circular, helicity-odd part), against 0.015 and 0.06 for linear light. Away from resonances, the $\alpha^2$ coefficient of $\slight$ is therefore formally correct for $\gamma\to0$; the finite-$\gamma$ calculation differs from it only by the weak $\gamma$ dependence of Sec.~S6.5, whereas the $\alpha^4$ part is not controlled. Fitting $\slight(\EF)/\alpha^2=c_2+c_4\alpha^2$ to $\alpha=0.005$--$0.02\angi$ gives $c_2=-3.85\ehA$ and $c_4=-2.3\times10^3\eh$\,\AA$^4$, i.e., $\alpha^4$ parts of 6\% at $\alpha=0.01\angi$ and 19\% at $0.02\angi$ of the fitted total (24\% of the quadratic term at $0.02\angi$); a fit to the two weakest fields alone gives the same $c_2$ ($-3.8500$ versus $-3.8497\ehA$). With $c_2$ alone, $\Delta^*$ at 0.1, 0.2 and 0.4\MVcm{} is 1.5, 6.1 and 29~meV, against 1.5, 6.5 and 38.6~meV with the full FFEH values; the main text quotes the leading-order values.

\textit{Integrated weak-field benchmark.} Because the helicity-odd curvature results from a near cancellation, we also compared the integrated coefficient $C(\alpha)=\slight(\EF)/\alpha^2$ with exact Floquet states at every point of identical $\bm k$ grids ($96^2$, $96^2$ shifted by half a spacing, and $144^2$; $\Delta=3$~meV, 50~K, dressed Fermi--Dirac occupations in both methods). The exact calculation diagonalizes the Floquet Hamiltonian with $N_F=3$, selects one family per band of $H^{(0)}$ by maximal overlap, and obtains the curvature from plaquette Berry phases together with the Wannier position connection. States more than $16k_BT$ below $\mu$ are treated as one subspace, whose time-sliced non-Abelian phase is insensitive to Pauli-blocked resonances among them; states within $16k_BT$ of $\mu$ are treated individually (grouping those within 2~meV, with their mean occupation, changes $C$ by $6\times10^{-5}$, relative). We regard two pointwise values $C_1$ and $C_2$ as converged when $|C_1-C_2|\le0.02\max(|C_2|,C_{\rm floor})$, a mixed relative and absolute tolerance whose floor $C_{\rm floor}=2.8$--$3.0\ehA$ is the median of $|C|$ in the FFEH and avoids a purely relative test near zeros of $C$. At 76--78\% of the points, $C$ is converged for $\alpha=0.00125$--$0.005\angi$ (plaquette side $10^{-4}$~\AA$^{-1}$). At the others, close to resonances where the perturbative range is narrower, we use $\alpha=3.1\times10^{-4}$ and $1.6\times10^{-4}\angi$ (side $5\times10^{-4}$~\AA$^{-1}$). For the 0.3--0.7\% of points that still fluctuate, we take the median over sides of $5\times10^{-4}$, $10^{-3}$ and $2\times10^{-3}$~\AA$^{-1}$, which removes artifacts of single plaquettes, such as a jump of the selected subspace across one link (link-overlap singular values down to $10^{-11}$); all but six points then converge, and these six contribute less than $10^{-4}$ of $C$. Including every grid point, the exact and FFEH values of $C$ agree within 0.7\% on all three grids ($-3.932$ and $-3.914$, $-3.862$ and $-3.835$, $-3.885$ and $-3.860\ehA$). On the $96^2$ grid, with the same per-point estimator for both cutoffs and every grid point included, $N_F=4$ changes the summed exact $C$ by at most $3\times10^{-5}$ (relative), provided that 54 points (0.6\%), which $N_F=4$ brings into new higher-order resonances at $\alpha=0.0025\angi$, are evaluated for both cutoffs at $\alpha=3.1\times10^{-4}$ or $1.6\times10^{-4}\angi$, where they return to their $N_F=3$ values; at $0.0025\angi$ these points alone shift the sum by 5.6\%, so agreement at most points does not by itself establish convergence of the integral. $N_F=2$ is insufficient. The FFEH values on the three grids lie within 1.7\% of $c_2=-3.85\ehA$ from the $1200^2$ grid, so the coarse grids represent the production integral; halving $\gamma$ changes the FFEH value by 1.2\% on the $96^2$ grid. A 1\% change of $c_2$ moves $\Delta^*$ at 0.4\MVcm{} by 0.3--0.4~meV. These comparisons test the weak-field coefficient, and the 0.7\% is the agreement between the two calculations, not an uncertainty of the predicted $\slight$, for which grid, regularization, occupation and higher-order effects are separate considerations. They do not test the $O(\alpha^4)$ terms or the regularized response near resonances at the fields of interest, where the exact $C$ depends on $\alpha$ through real multiphoton and weak exchange-activated transitions.

\subsection*{S6.5. Convergence}
\begin{figure*}[t]
\includegraphics[width=\textwidth]{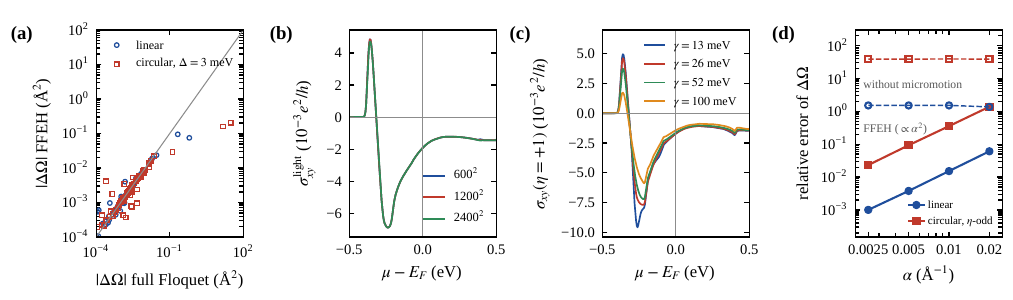}
\caption{FFEH at $\hbar\omega=0.2$~eV. (a) Pointwise light-induced Berry curvature of the Nb pair at 40 random $\bm k$ points ($\alpha=0.01\angi$): FFEH versus full Floquet ($N_F=8$), for linear light and for circular light with $\Delta=3$~meV; points far from the diagonal have another Floquet level within 1~meV. (b) $\sigma_{\rm odd}$ ($\alpha=0.02\angi$, $\Delta=3$~meV; Sec.~S9) for three grids. (c) $\sigma_{xy}(\eta=+1)$ for $\gamma=13$--100~meV ($1200^2$). (d) Check of the micromotion term: ratio of the median absolute error of the light-induced curvature to its median magnitude, FFEH versus exact Floquet states ($\gamma\to0$, Nb pair, 40 random $\bm k$), with (filled) and without (open) $\delta\bm A$, for linear light and for the helicity-odd part with circular light and $\Delta=3$~meV.}
\label{S:fig:ffeh}
\end{figure*}

The light part of the valve is converged on a $600^2$ grid (rms change to $2400^2$ below 1\%) [Fig.~\ref{S:fig:ffeh}(b)], and the BCD on $2400^2$ ($<1\%$ rms change to $4800^2$, peak 0.3\%). The light-induced part at $\Delta=3$~meV, $\sigma_{xy}(\EF;\eta=+1)-\sex(\EF)$, changes by $+0.1\%$, $-4.8\%$ and $-15\%$ for $\gamma=13$, 52 and 100~meV [$\sigma_{xy}(\EF;\eta=+1)$ itself, shown in Fig.~\ref{S:fig:ffeh}(c), by $+0.2\%$, $-5.3\%$ and $-17\%$]; the BCD peak by $+1.5\%$ and $-4.3\%$ for $\gamma=13$ and 52~meV (Sec.~S10).

\subsection*{S6.6. Validity map}
\begin{figure*}[t]
\includegraphics[width=0.66\textwidth]{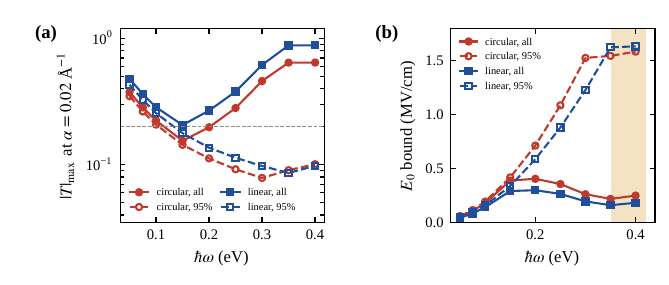}
\caption{Validity of the FFEH. (a) Largest Schrieffer--Wolff element $|T|$ for the states within $\EF\pm0.3$~eV ($144^2$ grid) for circular light with $\Delta=3$~meV and for linear light: maximum (filled) and 95th percentile (open), at $\alpha=0.02\angi$ (linear scaling from $\alpha=0.01\angi$); dashed: criterion $|T|=0.2$. (b) Resulting bound on $E_0$; beige: Nb--Se-$p$ near-resonance.}
\label{S:fig:valid}
\end{figure*}

We use as expansion parameter, for each state $n$, the largest band-basis element of the regularized generators, $|T|=\max_{l,p}|(T_p)_{ln}|$ with $(T_p)_{ln}=H^{(p)}_{ln}g(\varepsilon_n-\varepsilon_l-p\hbar\omega)$; it scales linearly with $\alpha$ to leading order (Bessel factors and dressed detunings add higher-order dependence). We evaluate it at $\alpha=0.01\angi$, rescaled linearly, with $p=\pm1$ and $\gamma=26$~meV, for all states within $\EF\pm0.3$~eV on a $\Gamma$-centred $144^2$ grid, for circular light with $\Delta=3$~meV (Hall valve) and for linear light without exchange (BCD). The $144^2$ grid is converged ($288^2$ and a shifted $144^2$ grid give the same maxima within 1\%), whereas a $48^2$ grid undersamples narrow regions near the Nb--Se-$p$ transitions and overestimates the bound at 0.2--0.3~eV (0.47 instead of 0.40\MVcm{} at 0.2~eV). Figure~\ref{S:fig:valid}(a) shows the maximum of $|T|$ and its 95th percentile; requiring $|T|\le0.2$ gives the bounds on $E_0$ in Fig.~\ref{S:fig:valid}(b) and main-text Fig.~4(a): for circular light 0.05, 0.18, 0.39, 0.40, 0.36 and 0.26\MVcm{} at $\hbar\omega=0.05$, 0.1, 0.15, 0.2, 0.25 and 0.3~eV, and for linear light, whose peak field is $\sqrt2$ larger at equal intensity, about 25\% lower (0.04, 0.14, 0.29, 0.30, 0.26 and 0.19\MVcm). For 95\% of the states the circular-light bounds are 0.19 and 1.09\MVcm{} at 0.1 and 0.25~eV. The circular-light frequency scan at $\alpha=0.02\angi$ therefore exceeds the bound by 12\% and 41\% at 0.1 and 0.25~eV and by factors of 1.9 and 2.3 at 0.05 and 0.3~eV; the linear-light scan exceeds it by factors of 1.4 and 3.1 at 0.1 and 0.3~eV and by 34\% at 0.2~eV, where the comparison with exact Floquet states below applies. At 0.2~eV the largest $|T|$ comes from a few states near M at $E-\EF\approx-0.3$~eV; for states within $\EF\pm0.05$~eV the bound is 0.68 (circular) and 0.53\MVcm{} (linear), and within $\pm0.05$~eV of the BCD peak 0.44 and 0.32\MVcm. For the same states at $\alpha=0.02\angi$ (circular light), the median $|T|$ and the median sideband weight $\sum_p\|T_p\psi_n\|^2$ are 0.25 and 13\% at 0.05~eV (not quantitative), 0.15 and 5\% at 0.1~eV, 0.08 and 1.4\% at 0.2~eV, and 0.05 and 0.7\% at 0.3~eV, where the maximum of $|T|$ reaches 0.46 at the Nb--Se-$p$ near-resonance; at 0.2~eV the largest sideband weight is 4.7\% (circular) and 8.0\% (linear). The criterion is a sampled diagnostic of the largest band-basis element, not an operator-norm bound. The regularized denominator is bounded by $|g|\le1/(2\gamma)$, with its maximum at $|x|=\gamma$, and vanishes at exact resonance; a small $|T|$ alone therefore cannot certify off-resonant validity, and resonant channels are examined separately (Sec.~S6.2). For linear light at 0.4\MVcm, where $|T|$ reaches 0.27, the comparison with exact Floquet states (Sec.~S6.4) gives a median absolute error of 6\% of the median light-induced curvature; we therefore use the BCD at 0.4\MVcm{} and show stronger fields only as trends.

\subsection*{S6.7. Relation to perturbative nonlinear-response theory}
The light-induced changes of $\sigma_{xy}$ and $\bm D$ are second order in the drive amplitude, so the currents they produce are third order (Hall current, $\propto E_{\rm drive}^2E_{\rm probe}$) and fourth order (BCD-mediated current, $\propto E_{\rm drive}^2E_{\rm probe}^2$) in the combined drive and probe fields. Perturbation theory in the drive gives the dressed-band terms as sums over virtual interband transitions with the denominators $\varepsilon_n-\varepsilon_l-p\hbar\omega$ of Eq.~\eqref{S:eq:ffeh}. For the Hall current, the light-dressed (Floquet) contribution has been identified explicitly within a third-order density-matrix calculation that also contains other mechanisms, such as the contribution induced by the bias field~\cite{S:Murotani2026}; the dressed-band terms computed here are therefore one part of a general third- or fourth-order response, not all of it. The FFEH collects these terms into the quasienergies, Berry curvatures and micromotion connection [Eq.~\eqref{S:eq:mm}] of dressed bands, so that both responses follow, as in the static problem, from Fermi-sea and Fermi-surface integrals over a band structure [Eq.~(3)], with the same Wannier interpolation and $k$ grids; this is the band-geometric form of light-induced responses of Ref.~\cite{S:Morimoto2016}. At order $\alpha^2$, the dressed-band contribution can be recovered from a consistent perturbative calculation with matching occupations, relaxation assumptions and probe limit. The integrated benchmark of Sec.~S6.4 tests the weak-field coefficient of the FFEH against exact Floquet states, not against such a calculation. The Floquet form makes three things explicit.

\textit{Mechanism.} The dressed bands locate the responses in the zone and in energy: the unequal shifts of the M saddles that unmask their hidden dipoles (Sec.~S8), and the near cancellation between the curvature of $\Heff$ and the micromotion term, which reverses the sign of $\slight$ relative to the high-frequency limit (Secs.~S6.4 and S9.4).

\textit{Occupations.} We compare inherited occupations, which represent negligible population redistribution during the dressing (as motivated by adiabatic switching without relaxation under nonresonant conditions), with the Fermi--Dirac occupations of the dressed bands used in the main text (Sec.~S5.4). Either prescription can also be implemented perturbatively: expanding $f(\varepsilon_n)$ about the undriven energies gives at order $\alpha^2$ the population term $(e^2/\hbar)\sum_n\int_{\bm k}(-\partial_\varepsilon f)\,\delta\varepsilon_n\Omega^0_n$ of Sec.~S5.4. Their difference isolates this population contribution, 27\% of $c_2$; removing it preserves the sign of $\slight$ and changes the leading-order threshold $\Delta^*$ at 0.4\MVcm{} from 29 to 20~meV. Which prescription applies depends on relaxation during the pulse (Sec.~S5.3).

\textit{Partial resummation of band shifts.} Because $\Heff$ is diagonalized and its quasienergies enter the Fermi functions unexpanded, $f(\varepsilon^0_n+\delta\varepsilon_n)$ retains selected higher powers of the light-induced shifts, such as the splitting of the M saddles (${\approx}2.5$~meV at $\alpha=0.01\angi$, growing as $\alpha^2$). This is useful near the BCD peak, where the splitting exceeds $k_BT$ at $\alpha=0.02\angi$ (${\approx}10$ against 4.3~meV at 50~K) and $D/\alpha^2$ becomes subquadratic (main text). It is a partial resummation within the truncated effective model, not a controlled higher-order result: terms of order $\alpha^4$ in the quasienergies and curvatures, and other processes of that order, are not all included, and a perturbative calculation could be reorganized to perform the same resummation. It also holds only within the model, whose Fermi surface is broadened by temperature only; a lifetime broadening $\hbar/\tau=26$~meV would move the onset to stronger fields.

Terms of order $\alpha^4$ in $\Heff$ and in the micromotion connection are not included (Sec.~S6.4), and the dissipative parts of the response (absorption, injection and shift currents), which perturbation theory contains, are absent from the Hermitian $\Heff$; they are suppressed below the dominant interband absorption onset (0.60~eV at $\EF$; Table~\ref{S:tab:checks}). The conclusions of the main text, including the micromotion-induced sign reversal of $\slight$, therefore concern the dressed-band physics rather than the choice of formalism.

% =====================================================================
\section*{S7. Selection rules in the mid-infrared}
\begin{figure*}[t]
\includegraphics[width=\textwidth]{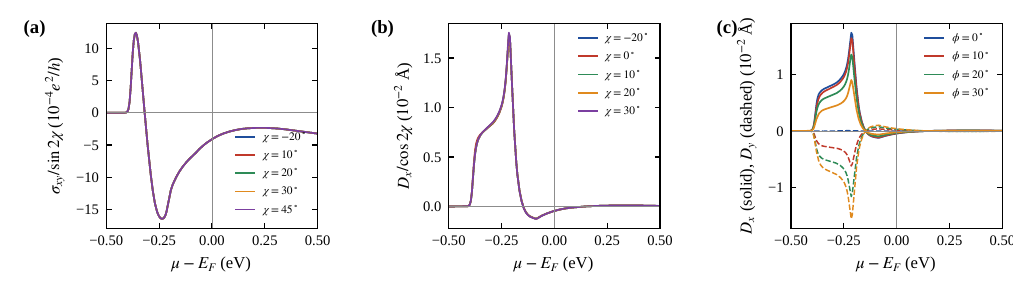}
\caption{Mid-infrared selection rules ($\hbar\omega=0.2$~eV, $\alpha=0.01\angi$). (a) $\sigma_{xy}/\sin2\chi$ and (b) $D_x/\cos2\chi$ for several ellipticities ($\phi=0$): the curves collapse. (c) $D_x$ (solid) and $D_y$ (dashed) for linear light at $\phi=0$--$30^\circ$.}
\label{S:fig:midIR}
\end{figure*}

% =====================================================================
Figure~\ref{S:fig:midIR} shows the full $\mu$ dependence behind main-text Fig.~1 ($\alpha=0.01\angi$). $\sigma_{xy}$ follows $\sin2\chi\,\sigma_{\rm circ}$ within 0.03\%, and $\bm D$ follows $\cos2\chi\,\bm D_{\rm lin}$ within 1.2--2.5\% (0.6--2.4\% at $\alpha=0.02\angi$), with $\bm D(\chi)=\bm D(-\chi)$ to numerical precision. The angle law $D_+=\beta e^{-2i\phi}$ holds within 4.7\% at $\alpha=0.01\angi$, with $|c_4/c_{-2}|=1.6\%$ at the peak; at $\alpha=0.02\angi$ the $C_3$-allowed $e^{4i\phi}$ harmonic grows to 5.5\% (fit residual 0.3\%), consistent with the onset of the subquadratic regime. Circular light gives $|\bm D|\le5.8\times10^{-5}$~\AA, the floor of the FFEH evaluation, which gives the same value at $\alpha=0$ (Sec.~S1).

% =====================================================================
\section*{S8. Mechanism of the light-induced BCD}
% =====================================================================
\subsection*{S8.1. Hidden dipoles at the M saddles}
The Nb pair has a van Hove saddle at M, $E_{\rm vH}-\EF=-0.2065$~eV, with second derivatives $\partial^2\varepsilon/\partial q^2=-8.4$~eV\,\AA$^2$ toward $\Gamma$ and $+43.9$~eV\,\AA$^2$ along the zone edge (average of the pair, which splits linearly along the edge). In the undriven crystal, a box of $\pm0.25\angi$ around each M point carries a large BCD: $D_x\approx-0.24$~\AA{} for M$_1$ (on the $k_y$ axis) and $+0.12$~\AA{} for M$_2$ and M$_3$, which cancel by $C_3$ (main-text Fig.~2(b)). The three M$_j$ are inequivalent modulo reciprocal lattice vectors; the six M points of the hexagonal zone in main-text Fig.~2(b) are their periodic images, and the arrows there integrate the half boxes inside the first zone, whereas the box sums quoted here integrate full boxes centred on M$_1$, M$_2$ and M$_3$, so that each saddle is counted once. We write the dipole of M$_j$, at angle $\theta_j$ (M$_1$: $\theta_1=90^\circ$), as $D_+^{(j)}=d(\mu)\,e^{i(\theta_j-\pi/2)}$, where $d(\mu)$ is the dipole of M$_1$ [Fig.~\ref{S:fig:mech}(a); $d=-0.233$~\AA{} and $\partial_\mu d=-3.05$~\AA/eV at $\mu-\EF=-0.215$~eV].

\subsection*{S8.2. Light-induced asymmetry}
Linear light shifts the saddles unequally: at $\alpha=0.01\angi$ and $\phi=0$, by $+2.95$~meV for M$_1$ (perpendicular to $\bm E$) and $+0.43$~meV for M$_2$ and M$_3$. The light-induced change of the BCD kernel is accordingly concentrated at M$_1$ [Fig.~\ref{S:fig:mech}(b); over the whole zone at $\alpha=0.02\angi$, the two M points on the $k_y$ axis carry 1.39 times the zone total, Fig.~\ref{S:fig:kmaps}(d)]: the three boxes contribute $(+2.14,\allowbreak\,-0.25,\allowbreak\,-0.25)\times10^{-2}$~\AA{}, in total $1.63\times10^{-2}$~\AA{} (sums formed before rounding), 94\% of the converged $1.732\times10^{-2}$~\AA.

\subsection*{S8.3. Saddle-point model}
\begin{figure*}[t]
\includegraphics[width=\textwidth]{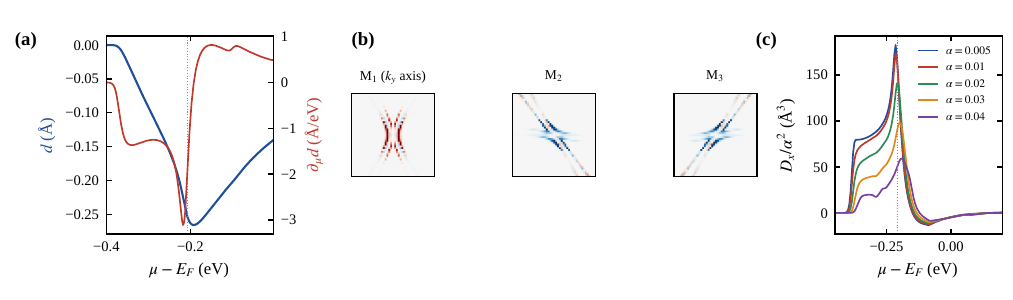}
\caption{BCD mechanism. (a) Static hidden dipole $d(\mu)$ of the box around M$_1$ and its derivative; dotted: saddle energy. (b) Light-induced change of the BCD kernel in $\pm0.25\angi$ boxes around M$_1$ ($k_y$ axis), M$_2$ and M$_3$ ($\alpha=0.01\angi$, $\phi=0$, 0.2~eV; common color scale, red positive). (c) $D_x/\alpha^2$ versus $\mu$ for $\alpha=0.005$--$0.04\angi$ (0.2~eV, $\phi=0$); dotted: saddle energy.}
\label{S:fig:mech}
\end{figure*}

Parametrizing the shifts as $\delta E_j=\delta_0+\delta_2\cos2(\phi-\theta_j)$ and the dressed dipoles as $d_j=d\,[1+\kappa_0+\kappa_2\cos2(\phi-\theta_j)]$,
\begin{equation}
\begin{split}
D_+=\sum_{j=1}^3{}&d\big(\mu-\delta E_j\big)\left[1+\kappa_0+\kappa_2\cos2(\phi-\theta_j)\right]\\
&\times e^{i(\theta_j-\pi/2)} .
\end{split}
\end{equation}
Since $\sum_je^{i\theta_j}=0$, $\sum_je^{-i\theta_j}=0$ and $e^{3i\theta_j}=-i$ for $\theta_j=90^\circ$, $210^\circ$, $330^\circ$, one finds $\sum_j\cos2(\phi-\theta_j)\,e^{i(\theta_j-\pi/2)}=-\tfrac32e^{-2i\phi}$, and to first order in the shifts and in $\kappa$
\begin{equation}
D_+=\tfrac32\left[\delta_2\,\partial_\mu d(\mu)-\kappa_2\,d(\mu)\right]e^{-2i\phi},
\end{equation}
which is Eq.~(5). The fitted parameters at $\alpha=0.01\angi$ are $\delta_0=+1.27$~meV, $\delta_2=-1.69$~meV, $\kappa_0=-0.028$ and $\kappa_2=+0.024$ ($\delta_0-\delta_2$ and $\delta_0+\delta_2/2$ reproduce the two shifts). The dipole-dressing term gives $0.84\times10^{-2}$~\AA{} (52\%). The energy-shift term gives $0.77\times10^{-2}$~\AA{} in the linearized form and $0.79\times10^{-2}$~\AA{} (48\%) when evaluated with the finite shifts, $D_x=d(\mu-\delta E_1)-d(\mu-\delta E_2)$, with $\delta E_1=2.953$~meV and $\delta E_2=0.425$~meV, both scaled as $(\alpha/0.01)^2$. This shift-only model gives $D_x/\alpha^2=79.1$, 78.9, 76.1, 68.4 and 58.3~\AA$^3$ at the peak for $\alpha=0.005$, 0.01, 0.02, 0.03 and $0.04\angi$ (squares in main-text Fig.~2(c)), against 182, 173, 141, 98 and 59~\AA$^3$ for the full calculation [Fig.~\ref{S:fig:mech}(c)]. The shift term alone reproduces the peak position and its drift with $\alpha$. The response becomes subquadratic once the saddle splitting $\delta E_1-\delta E_2$ (${\approx}2.5$~meV at $\alpha=0.01\angi$, growing as $\alpha^2$) exceeds $k_BT$; between $\alpha=0.005$ and $0.04\angi$ the shift-only $D/\alpha^2$ falls by 26\% and the full one by 68\%, so that the dressing term is largely quenched at the highest field. Thermal smearing of the same splitting accounts for the decrease with temperature [$(5.63,\allowbreak\,4.28,\allowbreak\,3.35)\times10^{-2}$~\AA{} at 50, 150 and 300~K].

\subsection*{S8.4. BCD at the Fermi level and in NbS$_2$}
The peak lies 0.2~eV below $\EF$. At $\EF$ the BCD is $D_x=-1.83\times10^{-3}$~\AA{} at $\alpha=0.02\angi$ and $-4.5\times10^{-4}$~\AA{} at $\alpha=0.01\angi$, opposite in sign and about 30 times smaller than the peak. Table~\ref{S:tab:doping} gives the carrier density needed to move $\mu$ in the free-standing monolayer (rigid band, $360^2$ grid, 50~K). Reaching the saddle requires $6.8\times10^{14}$ holes cm$^{-2}$ (0.69 holes per Nb, two thirds of the Nb-band filling), comparable to the doping at which the 30~meV proximity BCD of Ref.~\cite{S:Wicaksono2026} peaks ($5.6\times10^{14}$ electrons cm$^{-2}$ at $+0.28$~eV, $9.8\times10^{14}$ holes cm$^{-2}$ at $-0.37$~eV). Charge transfer of $10^{13}$~cm$^{-2}$ from a substrate shifts $\EF$ by only 4~meV, so, within the rigid-band approximation, the values at $\EF$ apply to weakly perturbing substrates in this charge-transfer range (hybridization and interface reconstruction are not included), whereas the peak requires deliberate doping. For states within $\EF\pm0.05$~eV the expansion parameter for linear light stays below 0.2 up to 0.53\MVcm{} (Sec.~S6.6).

\begin{table*}[t]
\caption{Carrier density $\delta n$ needed to shift the chemical potential of free-standing \NbSe{} (rigid band, 50~K; negative: holes) and light-induced $D_x$ at 0.2~eV and 0.4\MVcm{} (linear light, 50~K).}
\label{S:tab:doping}
\begin{ruledtabular}
\begin{tabular}{lcccccc}
$\mu-\EF$ (eV) & 0 & $-0.02$ & $-0.05$ & $-0.10$ & $-0.15$ & $-0.205$ \\ \hline
$\delta n$ ($10^{14}$~cm$^{-2}$) & 0 & $-0.54$ & $-1.40$ & $-3.21$ & $-4.85$ & $-6.79$ \\
$D_x$ ($10^{-2}$~\AA) & $-0.18$ & $-0.23$ & $-0.32$ & $-0.40$ & $+0.23$ & $+5.63$ \\
\end{tabular}
\end{ruledtabular}
\end{table*}

 In NbS$_2$ the BCD peak, at $\mu-\EF=-0.095$~eV, sits at its own M saddle ($-0.093$~eV relative to its DFT Fermi level), with $D/\alpha^2=20$~\AA$^3$, quadratic in $\alpha$ within numerical accuracy up to $0.02\angi$ (Sec.~S11); reaching it requires $2.9\times10^{14}$ holes cm$^{-2}$ (0.28 holes per Nb).

\subsection*{S8.5. Comparison with proximity-induced BCD}
The proximity values quoted in the main text are the peak magnitudes of $D_y$ in the same Wannier model of Ref.~\cite{S:Wicaksono2026} (WannierBerri, 50~K, $\mu-\EF\in[-0.5,0.5]$~eV, $D^{\rm 2D}=c\,D^{\rm 3D}$ with $c=30$~\AA): $1.07\times10^{-2}$~\AA{} for the two-sided orthogonal texture with $|\Delta_{\rm ex}|=30$~meV at both interfaces ($\mu-\EF=-0.370$~eV), $2.38\times10^{-2}$~\AA{} for the one-sided texture canted by $\theta=10^\circ$ ($+0.281$~eV), and $2.61\times10^{-2}$~\AA{} for the orthogonal texture with 30~meV out-of-plane and 5~meV in-plane exchange ($+0.280$~eV); the largest feature, $7.12\times10^{-2}$~\AA, occurs for 50~meV out-of-plane and 5~meV in-plane exchange ($+0.223$~eV), with a full width at half maximum of 23~meV. At 50~K the light-induced peak of $5.63\times10^{-2}$~\AA{} at $\alpha=0.02\angi$ therefore exceeds the peak magnitudes of the 30~meV configurations listed above; all these peaks lie away from $\EF$ and require doping of similar magnitude (Sec.~S8.4). At 300~K the light-induced peak is $3.35\times10^{-2}$~\AA; the proximity values were computed at 50~K only.

% =====================================================================
\section*{S9. Optical Hall valve: definitions and additional data}
% =====================================================================
\subsection*{S9.1. Definitions}
With $\sigma_\pm(\Delta)=\sigma_{xy}(\Delta,\eta=\pm1)$, we distinguish the helicity-odd response at finite exchange, $\sigma_{\rm odd}(\Delta)=[\sigma_+(\Delta)-\sigma_-(\Delta)]/2$, from its light-only value $\slight=\sigma_{\rm odd}(0)$, the light-induced conductivity of the main text. The exchange part $\sex(\Delta)$ is $\sigma_{xy}$ without light, $\sigma_{\rm even}(\Delta)=[\sigma_+(\Delta)+\sigma_-(\Delta)]/2$, and the cross terms are $\delta\sigma_{\rm even}=\sigma_{\rm even}-\sex$ and $\delta\sigma_{\rm odd}=\sigma_{\rm odd}-\slight$, so that exactly $\sigma_{xy}(\Delta,\eta)=\sex+\eta\slight+\delta\sigma_{\rm even}+\eta\,\delta\sigma_{\rm odd}$; the residual of the main text is $\delta\sigma(\Delta,\eta)=\delta\sigma_{\rm even}+\eta\,\delta\sigma_{\rm odd}$. In Fig.~\ref{S:fig:valve}(b,g), $\delta\sigma$ denotes $\delta\sigma_{\rm even}$. In practice we evaluate $\slight$ as $\sigma_{\rm odd}(3~{\rm meV})$, which agrees with the $\Delta=0$ value within 0.01\% at 0.2~eV and $\alpha=0.02\angi$ [$-1.90904\times10^{-3}\eh$ ($1200^2$) versus $-1.90886\times10^{-3}\eh$ ($2400^2$)]; figure panels and table entries specified with $\Delta=3$~meV show $\sigma_{\rm odd}(3~{\rm meV})$, labelled $\sigma_{xy}^{\rm light}$ on the figure axes. The sign-reversal fraction is the share of the 201 $\mu$ points in $\mu-\EF\in[-0.5,0.5]$~eV where $\sigma_{xy}(\eta=+1)$ and $\sigma_{xy}(\eta=-1)$ have opposite signs and both exceed $1\times10^{-4}\eh$ in magnitude (fixed threshold, the same for all $\Delta$). The helicity reverses the sign of $\sigma_{xy}(\EF)$ exactly when $|\sigma_{\rm odd}|>|\sigma_{\rm even}|$; because both cross terms are small at $\EF$ (Table~\ref{S:tab:valve}), this reduces approximately to the additive criterion $|\slight(\EF)|>|\sex(\EF)|$ (the two thresholds coincide only if the cross terms vanish; the absolute value matters for NbTe$_2$, where $\sex<0$). The exchange $\Delta$ multiplies the Pauli matrix $s_z$, so a pure-spin orbital is split by $2\Delta$ (the Nb pair at M by $\pm5$~meV at $\Delta=30$~meV). The crossover exchange $\Delta^*$ is defined at $\EF$ by $|\sex(\EF;\Delta^*)|=|\slight(\EF)|$, interpolating $\sex(\EF;\Delta)$ between the computed exchanges (1--120~meV) and taking the first crossing.

\begin{table*}[t]
\caption{Hall valve at $\EF$ versus exchange ($\hbar\omega=0.2$~eV, $\alpha=0.02\angi$, 50~K). Units $10^{-3}\,e^2/h$ unless stated.}
\label{S:tab:valve}
\begin{ruledtabular}
\begin{tabular}{lcccccc}
$\Delta$ (meV) & 1 & 3 & 10 & 30 & 40 & 50 \\ \hline
$\sigma_{xy}(\eta=+1)$ & $-1.842$ & $-1.711$ & $-1.299$ & $-0.323$ & $+0.043$ & $+0.343$ \\
$\sigma_{xy}(\eta=-1)$ & $+1.976$ & $+2.107$ & $+2.528$ & $+3.540$ & $+3.920$ & $+4.223$ \\
$\sigma_{\rm odd}$ (light only, $\slight$: $-1.909$) & $-1.909$ & $-1.909$ & $-1.914$ & $-1.931$ & $-1.938$ & $-1.940$ \\
$\delta\sigma_{\rm odd}$ & $0.000$ & $0.000$ & $-0.005$ & $-0.022$ & $-0.029$ & $-0.031$ \\
$\delta\sigma_{\rm even}$ & $0.0005$ & $0.0009$ & $-0.0047$ & $0.0134$ & $0.0216$ & $0.0349$ \\
$\sex$ & 0.0663 & 0.1974 & 0.6194 & 1.595 & 1.960 & 2.248 \\
$\sex/\Delta$ ($10^{-5}\,e^2/h$ per meV) & 6.63 & 6.58 & 6.19 & 5.32 & 4.90 & 4.50 \\
Helicity reverses $\sigma_{xy}(\EF)$ & yes & yes & yes & yes & no & no \\
Sign-reversal fraction & 89\% & 88\% & 85\% & 66\% & 46\% & 36\% \\
\end{tabular}
\end{ruledtabular}
\end{table*}

\subsection*{S9.2. Additivity, sign-control crossover and field dependence}
\begin{figure*}[t]
\includegraphics[width=\textwidth]{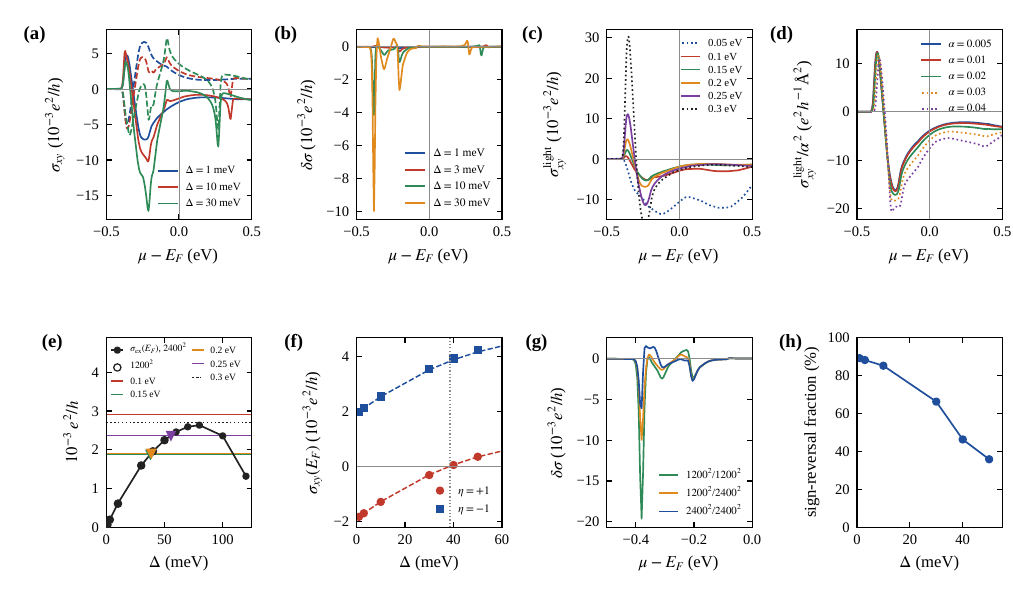}
\caption{Hall valve, additional data ($\hbar\omega=0.2$~eV, 50~K). (a) $\sigma_{xy}$ for $\Delta=1$, 10 and 30~meV with $\eta=+1$ (solid) and $\eta=-1$ (dashed), $\alpha=0.02\angi$. (b) Helicity-even cross term $\delta\sigma_{\rm even}$ for $\Delta=1$--30~meV. (c) $\sigma_{\rm odd}$ for $\hbar\omega=0.05$--0.3~eV at $\alpha=0.02\angi$ ($\Delta=3$~meV; dotted: trends only). (d) $\sigma_{\rm odd}/\alpha^2$ for $\alpha=0.005$--$0.04\angi$ ($\Delta=3$~meV; dotted: beyond the validity bound). (e) Exchange-only $\sex(\EF)$ versus $\Delta$ (filled: $2400^2$; open: $1200^2$) and $|\sigma_{\rm odd}(\EF)|$ ($\Delta=3$~meV) at $\alpha=0.02\angi$ for $\hbar\omega=0.1$--0.3~eV (horizontal lines; dotted: trends only); triangles: $\Delta^*$ (FFEH). (f) $\sigma_{xy}(\EF)$ for $\eta=\pm1$ from full calculations (symbols) and the additive estimate $\sex+\eta\slight$ (dashed), $\alpha=0.02\angi$; dotted: $\Delta^*=38.6$~meV (full FFEH value; 28.9~meV at leading order). (g) Helicity-even cross term $\delta\sigma_{\rm even}$ at $\Delta=30$~meV for three combinations of driven/static grids. (h) Sign-reversal fraction versus $\Delta$ ($\alpha=0.02\angi$).}
\label{S:fig:valve}
\end{figure*}

At $\EF$ the valve is additive up to the largest exchange of the full calculations (Table~\ref{S:tab:valve}): for $\Delta\le50$~meV, $|\delta\sigma_{\rm odd}|\le3.1\times10^{-5}\eh$ (1.6\% of $\slight$) and $|\delta\sigma_{\rm even}|\le3.5\times10^{-5}\eh$; the additive estimate, which neglects both, reproduces the full calculations within $7\times10^{-5}\eh$. Away from $\EF$ the helicity-even cross term is larger [Fig.~\ref{S:fig:valve}(b)]: near the saddle ($-0.205$~eV) it reaches 13\% ($\Delta=10$~meV) and 32--39\% ($\Delta=30$~meV, depending on the grid) of the light-induced peak, and 30--60\% at 40--50~meV. At the band edge ($-0.38$~eV) the cross term is not converged with the $k$ grid: for $\Delta=30$~meV its maximum falls from 275\% to 140\% and 86\% of the light-induced peak for the $1200^2/1200^2$, $1200^2/2400^2$ and $2400^2/2400^2$ combinations of driven and static grids [Fig.~\ref{S:fig:valve}(g)], reflecting a sharp band-edge feature that these grids do not resolve. At $\EF$, $\delta\sigma_{\rm even}$ is grid independent ($(1.34$--$1.35)\times10^{-5}\eh$ at 30~meV), and $\sex(\EF)$ changes by at most 1.6\% between $1200^2$ and $2400^2$.

$\sex(\EF)$ is sublinear in $\Delta$ and not monotonic: $\sex(\EF)=(1.960,\allowbreak\,2.248,\allowbreak\,2.461,\allowbreak\,2.593,\allowbreak\,2.635,\allowbreak\,2.359,\allowbreak\,1.318)\times10^{-3}\eh$ for $\Delta=40$, 50, 60, 70, 80, 100 and 120~meV, with a maximum near 80~meV [Fig.~\ref{S:fig:valve}(e)]. At leading order in the field (Sec.~S6), $\slight(\EF)=c_2\alpha^2$ with $c_2=-3.85\ehA$ (fit of $\slight/\alpha^2=c_2+c_4\alpha^2$ to $\alpha=0.005$--$0.02\angi$, $c_4=-2.3\times10^3\eh$\,\AA$^4$), and interpolation within the computed range gives $\Delta^*=1.5$, 6.1 and 28.9~meV for $E_0=0.1$, 0.2 and 0.4\MVcm{} ($\alpha=0.005$, 0.01 and $0.02\angi$); these are the values of the main text. With the full FFEH values of $\slight(\EF)$, which contain part of the $\alpha^4$ terms, $\Delta^*=1.5$, 6.5 and 38.6~meV, and the full calculations bracket the last value [Fig.~\ref{S:fig:valve}(f)]: the helicity reverses $\sigma_{xy}(\EF)$ at 30~meV [$(-0.32,\allowbreak\,+3.54)\times10^{-3}\eh$ for $\eta=\pm1$] but not at 40~meV [$(+0.04,\allowbreak\,+3.92)\times10^{-3}\eh$] or 50~meV [$(+0.34,\allowbreak\,+4.22)\times10^{-3}\eh$]; linear interpolation of $\sigma_{xy}(\EF;\eta=+1)$ between 30 and 40~meV puts its zero at 38.8~meV. Above the first crossover, the exchange sets the sign over the following same-sign interval. Because $\sex(\EF)$ is nonmonotonic, the additive estimate predicts a second crossing, i.e., a re-entrant sign reversal, at about 116~meV at 0.4\MVcm{} at leading order (109~meV with the full FFEH value), beyond the range in which additivity has been tested (Sec.~S9.3). The ratio $r=\sigma_{xy}(\EF;\eta=+1)/\sigma_{xy}(\EF;\eta=-1)$ of the additive estimate is largest where $\sex(\EF)$ is largest, near 80~meV: $r_{\max}=0.26$, 0.75 and 0.93 at 0.4, 0.2 and 0.1\MVcm{} at leading order (0.16, 0.73 and 0.93 with the full FFEH values). In the same-sign regime the leading-order additive estimate thus gives $0<r\le0.26$ at 0.4\MVcm, a magnitude contrast of at least 3.8 (main-text Fig.~3); where $r<0$, the helicity instead reverses the sign, and the magnitude contrast need not exceed 3.8 (1.07 at $\Delta=1$~meV, Table~\ref{S:tab:valve}). The additive estimate reproduces the full calculations with exchange ($\Delta=1$--50~meV at 0.4\MVcm, 3~meV at 0.1 and 0.2\MVcm, both helicities) within $7\times10^{-5}\eh$; beyond 50~meV it is an extrapolation. At 0.6\MVcm{} ($\alpha=0.03\angi$, beyond the validity bound), $|\slight(\EF)|=5.3\times10^{-3}\eh$ exceeds the maximum of $\sex(\EF)$, so that no crossover occurs below 120~meV. In the FFEH the light part grows faster than $\alpha^2$ beyond $\alpha=0.01\angi$: $\slight(\EF)/\alpha^2=-3.9$, $-4.1$, $-4.8$, $-5.9$ and $-7.6\ehA$ for $\alpha=0.005$, 0.01, 0.02, 0.03 and $0.04\angi$ [Fig.~\ref{S:fig:valve}(d); the last two beyond the validity bound]. The sign-reversal fraction is 89, 88, 85, 66, 46 and 36\% for $\Delta=1$, 3, 10, 30, 40 and 50~meV [Fig.~\ref{S:fig:valve}(h)].

\subsection*{S9.3. Frequency and temperature dependence}
At $\alpha=0.02\angi$, $\slight(\EF)=-1.05\times10^{-2}\eh$ at 0.05~eV (not quantitative) and $(-2.9,\allowbreak\,-1.9,\allowbreak\,-1.9,\allowbreak\,-2.4,\allowbreak\,-2.7)\times10^{-3}\eh$ at $\hbar\omega=0.1$, 0.15, 0.2, 0.25 and 0.3~eV (0.3~eV: trend only) [Fig.~\ref{S:fig:valve}(c)], against $+4\times10^{-4}\eh$ at 10~eV (full Floquet). The sign-reversal fraction is 88.6, 87.6, 88.1, 88.1, 89.1 and 90.0\% at these frequencies ($\Delta=3$~meV). With $\sex(\EF)$ known up to 120~meV, the full FFEH values give $\Delta^*=37.9$, 38.6 and 55.5~meV at 0.15, 0.2 and 0.25~eV for $\alpha=0.02\angi$ [Fig.~\ref{S:fig:valve}(e)]; field scans, needed for the leading-order values, exist only at 0.2~eV (28.9~meV). At 0.1 and 0.3~eV, $|\slight(\EF)|=2.9\times10^{-3}$ and $2.7\times10^{-3}\eh$ exceeds the maximum of $\sex(\EF)$, $2.64\times10^{-3}\eh$ at $\Delta=80$~meV, so that the helicity sets the sign for every computed exchange; at $\alpha=0.02\angi$, however, the drive exceeds the circular-light bound by 12\% at 0.1~eV and by a factor of 2.3 at 0.3~eV (trend only), and the same holds at 0.05~eV, where the result is not quantitative (Sec.~S6.6). Because $\sex(\EF)$ decreases beyond 80~meV, the additive estimate also predicts a re-entrant valve above ${\approx}100$--110~meV at 0.15--0.25~eV; this lies beyond typical proximity exchange and beyond the range ($\le50$~meV) in which additivity has been checked. With temperature, $\slight(\EF)=(-1.91,\allowbreak\,-1.93,\allowbreak\,-1.99)\times10^{-3}\eh$ at 50, 150 and 300~K, with sign-reversal fractions of 88, 91.5 and 93.5\% [Fig.~\ref{S:fig:robust}(c)], and $(-2.00,\allowbreak\,-2.12,\allowbreak\,-2.13,\allowbreak\,-2.14)\times10^{-3}\eh$ at 335, 520, 540 and 550~K. For temperature-independent bands, curvatures and regularization, as in the FFEH, responses at two temperatures obey the exact identity $Q(\mu,T_2)=\int d\mu'\,Q(\mu',T_1)K(\mu-\mu')$ for $Q=\sigma_{xy}$ or $\bm D$, with Fourier transform $\tilde K(q)=(T_2/T_1)\sinh(\pi k_BT_1q)/\sinh(\pi k_BT_2q)$ ($T_2>T_1$). We evaluate it on the 5~meV grid of the computed 50~K ($\sigma_{xy}$) and 300~K ($\bm D$) curves, continued beyond the window $\mu-\EF\in[-0.5,0.5]$~eV by a cosine taper. This reproduces the direct 150 and 300~K calculations to $10^{-5}$ ($\slight$) and $10^{-6}$ (BCD peak), relative. At 550~K, replacing the taper by zero or constant continuation over 0.1--0.6~eV, changing the zero padding fourfold, or starting from 50, 150 or 300~K changes $\slight(\EF)$ by at most $2\times10^{-5}$ and the BCD peak by at most $1.4\times10^{-4}$, relative. The quoted BCD peaks are maxima over $\mu$ at each temperature, located at $\mu-\EF=-0.205$, $-0.230$ and $-0.250$~eV at 50, 300 and 550~K. Along a fixed-density heating trajectory (undriven bands, $600^2$ grid), a sample at the density of the 150~K peak has $D_x=3.15$ and $2.50\times10^{-2}$~\AA{} at 335 and 540~K, 1.5\% and 2\% below the maxima, and for the undoped sample $\slight(\EF)$ changes by 2.6\% at 300~K and 7.5\% at 540~K, less than at fixed $\mu$. Rescaling $\sex(\Delta)$ in the same way, the leading-order $\Delta^*$ at 0.1, 0.2 and 0.4\MVcm{} is 1.5, 6.2 and 29.2~meV at 300~K and 1.5, 6.3 and 30.0~meV at 335~K. At 520 and 540~K it rises to 1.9 and 2.0~meV at 0.1\MVcm{} and to 8.1 and 8.4~meV at 0.2\MVcm, and at 0.4\MVcm{} $|\slight(\EF)|$ exceeds $\sex(\EF)$ for all computed exchange (maximum $1.6\times10^{-3}\eh$ at 50--60~meV), which beyond 50~meV relies on the additive form. Electron heating therefore does not appreciably narrow the leading-order valve window in the additive estimate.

\subsection*{S9.4. Origin of the sign change relative to the high-frequency limit}
\begin{figure*}[t]
\includegraphics[width=\textwidth]{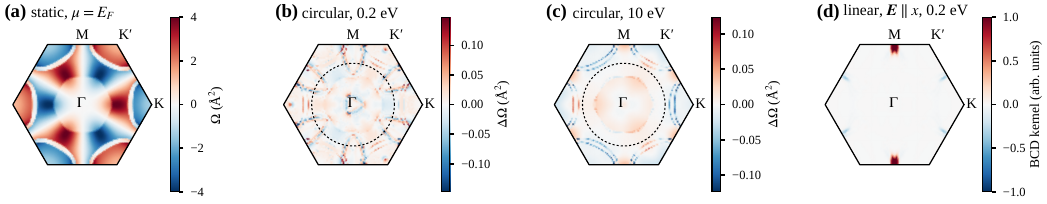}
\caption{$k$-resolved maps for \NbSe{} without exchange ($\EF=-2.4083$~eV; $110\times110$ Cartesian grid, 7004 points in the first Brillouin zone, spacing $0.0236\angi$; Gouraud-shaded Delaunay triangulation; each panel has its own symmetric color scale, set by the 99.5th percentile of the magnitude; M is on the $k_y$ axis). (a) Occupied Berry curvature $\sum_nf(\varepsilon_n-\EF)\Omega_n$ of the undriven model at 50~K. (b) Light-induced change of the occupied Berry curvature, $\delta\mathcal B=\sum_nf^{\rm drv}_n\Omega^{\rm drv}_n-\sum_nf^0_n\Omega^0_n$ (each with its own occupations), for circular light ($\eta=+1$) at $\hbar\omega=0.2$~eV and $\alpha=0.02\angi$ ($E_0=0.4$\MVcm), FFEH, 50~K. (c) Same at $\hbar\omega=10$~eV ($E_0=20$\MVcm), computed with the FFEH (not the first-order $\Heff$). Dashed circles: boundary of the $\Gamma$ region used for the region sums in Sec.~S9. (d) Light-induced change of the BCD kernel $\sum_nv_{x,n}\Omega_n(-\partial f/\partial\varepsilon)$ at $\mu-\EF=-0.210$~eV for linear light along $x$ ($\hbar\omega=0.2$~eV, $\alpha=0.02\angi$), evaluated with 300~K Fermi smoothing because the grid cannot resolve the 50~K Fermi window (arbitrary units, normalized to the 99.5th percentile). The two M points on the $k_y$ axis, perpendicular to $\bm E$, carry 1.39 times the zone total; the other four partly cancel.}
\label{S:fig:kmaps}
\end{figure*}

Maps of the light-induced Berry curvature over the Brillouin zone [Fig.~\ref{S:fig:kmaps}(b,c)] locate the sign change. They are computed with the FFEH for light alone ($\Delta=0$, $\eta=+1$, $\alpha=0.02\angi$) on a $110\times110$ Cartesian grid, of which 7004 points (spacing $0.0236\angi$) lie in the first Brillouin zone, and are therefore qualitative. Summed over the zone, the change of the occupied Berry curvature, $\delta\mathcal B(\bm k)=\sum_nf_n^{\rm drv}\Omega_n^{\rm drv}-\sum_nf_n^0\Omega_n^0$, with Fermi--Dirac occupations of the driven and undriven bands, respectively, gives raw grid sums (in \AA$^2$) that split into a $\Gamma$ region ($|\bm k|<0.73\angi\approx0.6|{\rm K}|$, dashed circles in the figure) and the K/K$'$ region (the rest) as $+3.16$ and $+17.45$~\AA$^2$ at 0.2~eV (total $+20.61$~\AA$^2$) and as $+15.65$ and $-21.77$~\AA$^2$ at 10~eV (total $-6.12$~\AA$^2$); a positive total means $\slight<0$. Converted with $\sigma_{xy}/(e^2/h)=-(2\pi/AN_k)\sum_{\bm k}\delta\mathcal B(\bm k)$ ($A=10.14$~\AA$^2$, $N_k=7004$), the totals give $\slight(\EF)\approx-1.8\times10^{-3}$ and $+5\times10^{-4}\eh$, consistent in sign and magnitude with the converged values ($-1.91\times10^{-3}$ and $+4\times10^{-4}\eh$). The $\Gamma$ region contributes with the same sign at both frequencies; the reversal comes from the K/K$'$ pockets, which dominate at 0.2~eV and outweigh the $\Gamma$ region with the opposite sign at 10~eV. At 10~eV the FFEH is used rather than the first-order $\Heff$, whose circular-light Hall response there is about 1.5 times the full Floquet value (Sec.~S5). A partner-resolved test on the same grid and in the same units, in which the intermediate bands $l$ of $H^{(0)}$ are masked in $\Heff$ and in the micromotion generators, is nearly additive at 10~eV: the $H^{(0)}$ renormalization alone gives $+6.38$~\AA$^2$, and the Se~$p$, Nb-pair and upper-$d$ partners add $-4.55$, $-5.23$ and $-2.28$~\AA$^2$, a sum of $-5.68$~\AA$^2$ against $-6.12$~\AA$^2$ for the full calculation. At 0.2~eV the decomposition is not meaningful: the parts sum to $-321$~\AA$^2$ (Nb pair alone $-306$~\AA$^2$) against a total of $+20.6$~\AA$^2$. Intraband $1/\hbar\omega$ terms, contained in the Nb-pair group, cancel against interband ones through sum rules, so single partners are not gauge meaningful in the velocity gauge. An interpretation in terms of red versus blue detuning from the Nb--Se-$p$ transitions (0.35--0.8~eV) therefore remains tentative; a length-gauge decomposition would be needed.

% =====================================================================
\section*{S10. Robustness}
\begin{figure*}[t]
\includegraphics[width=\textwidth]{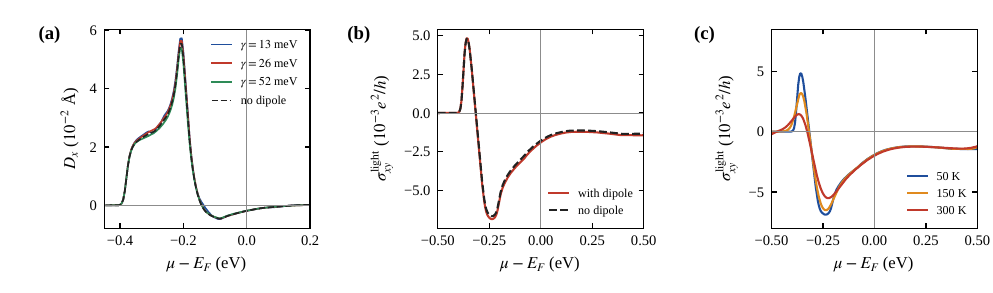}
\caption{Robustness ($\hbar\omega=0.2$~eV, $\alpha=0.02\angi$, 50~K). (a) $D_x$ for $\gamma=13$, 26 and 52~meV and without the dipole coupling. (b) $\sigma_{\rm odd}$ ($\Delta=3$~meV) with and without the dipole coupling. (c) $\sigma_{\rm odd}$ ($\Delta=3$~meV) at electron temperatures of 50, 150 and 300~K.}
\label{S:fig:robust}
\end{figure*}

% =====================================================================
\begin{table*}[t]
\caption{Sensitivity of the main results ($\hbar\omega=0.2$~eV, $\alpha=0.02\angi$ unless stated). Percentages are signed relative changes, $X/X_{\rm ref}-1$; since no quantity changes sign, they are also the relative changes of $|X|$. Entries with $<$ or $\le$ are bounds on $|X/X_{\rm ref}-1|$. The last column is $\sigma_{\rm odd}(\EF)$ at $\Delta=3$~meV; the $\gamma$ entries use $\sigma_{xy}(\eta=+1)-\sex$, which differs from it by $\delta\sigma_{\rm even}\approx10^{-6}\eh$.}
\label{S:tab:robust}
\begin{ruledtabular}
\begin{tabular}{p{0.30\textwidth}p{0.30\textwidth}p{0.34\textwidth}}
Variation & BCD peak & $\sigma_{\rm odd}(\EF)$, $\Delta=3$~meV \\ \hline
$\gamma=13$ / 52~meV (reference 26~meV) & $+1.5\%$ / $-4.3\%$ & $+0.1\%$ / $-4.8\%$ ($\gamma=100$~meV: $-15\%$) \\
Dipole coupling off & $-1.7\%$ & $-6.0\%$ \\
$T=150$ / 300 / 550~K & $(4.28,\allowbreak\,3.35,\allowbreak\,2.52)\times10^{-2}$~\AA{} (50~K: $5.63\times10^{-2}$~\AA) & $(-1.93,\allowbreak\,-1.99,\allowbreak\,-2.14)\times10^{-3}\eh$ (50~K: $-1.91\times10^{-3}\eh$) \\
$k$ grid & $2400^2$ vs $4800^2$: $<1\%$ & $600^2$ vs $2400^2$: $<1\%$ \\
$\EF$ re-referencing ($-5.9$~meV) & --- & $\le3\%$ \\
Grouping threshold 0.5 / 1 / 4~meV (ref.\ 2) & $+0.9$ / $+0.9$ / $-1.7\%$ & $\le0.01\%$ \\
\end{tabular}
\end{ruledtabular}
\end{table*}
Table~\ref{S:tab:robust} and Fig.~\ref{S:fig:robust} summarize the sensitivity. The value at $\EF$ of the BCD stays within 3\% for the broadening variations, and the mirror-forbidden $D_y$ remains at the floor. The dipole coupling, proportional to $\hbar\omega\alpha$, is a small correction in the mid-infrared.

% =====================================================================
\section*{S11. N\lowercase{b}S$_2$ and N\lowercase{b}T\lowercase{e}$_2$}
\begin{figure*}[t]
\includegraphics[width=\textwidth]{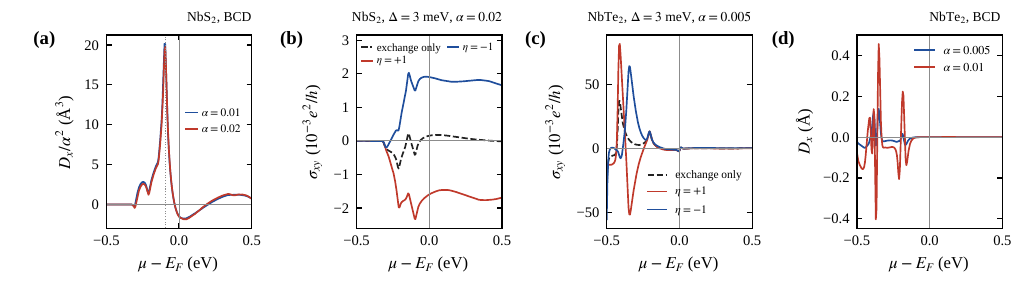}
\caption{NbS$_2$ and NbTe$_2$ at $\hbar\omega=0.2$~eV (50~K). (a) NbS$_2$: $D_x/\alpha^2$ for linear light; dotted: M saddle. (b) NbS$_2$: $\sigma_{xy}$ for $\Delta=3$~meV without light and with $\eta=\pm1$ ($\alpha=0.02\angi$). (c) NbTe$_2$: same for $\alpha=0.005\angi$. (d) NbTe$_2$: $D_x$ for linear light (outside the quantitative regime).}
\label{S:fig:mat}
\end{figure*}

% =====================================================================
The chalcogen-$p$ to Nb gap is 0.61~eV in NbS$_2$, a cleaner window than in \NbSe, and 0.08~eV in NbTe$_2$, which is near resonant in the mid-infrared (validity limit 0.12\MVcm{} at 0.2~eV). The M saddle lies at $-0.093$~eV in NbS$_2$ and at $-0.176$~eV in NbTe$_2$, relative to their DFT Fermi levels (Fig.~\ref{S:fig:bands}). The Nb pair of NbTe$_2$ carries more chalcogen weight: its exchange splitting at M is $\pm13$~meV at $\Delta=30$~meV, against $\pm5$~meV in NbS$_2$ and \NbSe. The symmetry floors hold for both compounds. The light part of the valve per $\alpha^2$ is similar across the three compounds: $-3.45$ and $-4.37\ehA$ for NbS$_2$ ($\alpha=0.01$, $0.02\angi$), $-4.1$ and $-4.8$ for \NbSe, and $-3.71$ and $-3.83$ for NbTe$_2$ ($\alpha=0.005$, $0.01\angi$; the latter beyond its validity limit), i.e., leading-order coefficients $c_2=-3.15$, $-3.85$ and $-3.67\ehA$. The exchange response per meV at $\EF$ differs: $+5.1\times10^{-5}$ (NbS$_2$), $+6.6\times10^{-5}$ (\NbSe{} at 1~meV, sublinear beyond) and $-4.7\times10^{-4}\eh$ (NbTe$_2$ at 3~meV). In NbS$_2$, $\sex(\EF)$ stays linear in $\Delta$: the response per meV is constant within 3\% up to 100~meV [$\sex(\EF)=(0.512,\allowbreak\,1.545,\allowbreak\,2.579,\allowbreak\,4.082,\allowbreak\,4.976)\times10^{-3}\eh$ at 10, 30, 50, 80 and 100~meV]. NbS$_2$ is therefore also a valve [Fig.~\ref{S:fig:mat}(b)]: at leading order ($c_2=-3.15\ehA$ from $\alpha=0.01$ and $0.02\angi$, with $\alpha^4$ parts of 9 and 28\%), $\Delta^*=6.1$ and 24.5~meV at $\alpha=0.01$ and $0.02\angi$; the full FFEH values give 6.7 and 33.9~meV, and full calculations at $\alpha=0.02\angi$ bracket the latter. The helicity reverses $\sigma_{xy}(\EF)$ at 30~meV [$(-0.21,\allowbreak\,+3.29)\times10^{-3}\eh$ for $\eta=\pm1$] but not at 40~meV [$(+0.30,\allowbreak\,+3.80)\times10^{-3}\eh$], with the zero of $\sigma_{xy}(\EF;\eta=+1)$ at 34.2~meV; up to 60~meV the helicity-odd part stays within 0.5\% of the light-only value and $|\delta\sigma_{\rm even}(\EF)|\le2.3\times10^{-5}\eh$. In NbTe$_2$, $\sex(\EF)$ is not linear in $\Delta$: a sharp feature within about 0.1~eV of $\EF$ gives $-2.5\times10^{-4}\eh$ at 1~meV and $-1.41\times10^{-3}\eh$ at 3~meV, whereas away from $\EF$ the response scales linearly. Interpolation gives $\Delta^*\approx0.4$ and 1.2~meV at $\alpha=0.005$ and $0.01\angi$, at leading order ($c_2=-3.67\ehA$) as with the full values, so in NbTe$_2$ the exchange sets the sign for exchange exceeding these crossover values under these drive conditions [Fig.~\ref{S:fig:mat}(c)]. The NbS$_2$ BCD peaks at its saddle with $D/\alpha^2=20$~\AA$^3$, about nine times smaller than in \NbSe{} [Fig.~\ref{S:fig:mat}(a)]. The NbTe$_2$ BCD is very large (0.14--0.46~\AA) at $-0.35$~eV, in the near-resonant Te-$p$ region, and already nonlinear at $\alpha=0.005\angi$; it is not quantitative [Fig.~\ref{S:fig:mat}(d)].

% =====================================================================
\section*{S12. Heating, working window, charge-density wave and detection}
% =====================================================================
\subsection*{S12.1. Drude heating}
Intraband absorption of a sheet with dc conductance $\sigma_{\rm dc}$ is $P=\sigma_{\rm dc}E_0^2/(1+\omega^2\tau^2)$. At fixed $\alpha$, $E_0=\hbar\omega\alpha/e$ and $P\to\sigma_{\rm dc}(\hbar\alpha/e\tau)^2$ for $\omega\tau\gg1$, nearly independent of $\hbar\omega$. With $\tau=25$~fs, $\sigma_{\rm dc}=1$--5~mS and $\alpha=0.02\angi$, $P\approx2.7$--14~MW/cm$^2$. For a pulse of length $t_p$, an electronic heat capacity $C_e=\gamma_eT_e$ gives $T_e=(T_0^2+2Pt_p/\gamma_e)^{1/2}$ (constant $\gamma_e$). Both heat capacities are assumed inputs. We use $\gamma_e=3.0\times10^{-7}$~J\,m$^{-2}$K$^{-2}$, which includes a mass enhancement: it is 2.9 times the bare band value of our model, $\gamma_e^0=\pi^2k_B^2N(\EF)/3A=1.0\times10^{-7}$~J\,m$^{-2}$K$^{-2}$ [$N(\EF)=2.64$ states/eV per formula unit, both spins]. The enhancement weakens once $k_BT_e$ exceeds the phonon energies. The two assumed values illustrate the sensitivity of the estimated $T_e$. For the lattice we use the Dulong--Petit value $C_{\rm lat}=9k_B/A=1.2\times10^{-3}$~J\,m$^{-2}$K$^{-1}$; transferring all absorbed energy to the lattice then gives $\Delta T_{\rm lat}=Pt_p/C_{\rm lat}$. From $T_0=50$~K and $\sigma_{\rm dc}=5$~mS, a 100~fs pulse yields $T_e\approx158$~K at $\alpha=0.01\angi$ and $\approx304$~K at $\alpha=0.02\angi$ ($\approx262$ and $\approx517$~K with $\gamma_e^0$); from $T_0=150$~K, above $T_{\rm CDW}$, the same absorbed energy gives $T_e\approx212$ and $\approx335$~K ($\approx298$ and $\approx536$~K with $\gamma_e^0$). For 1--5~mS at $\alpha=0.02\angi$, $T_e\approx140$--310~K ($235$--520~K with $\gamma_e^0$) and $\Delta T_{\rm lat}\approx2$--11~K for 100~fs, and $T_e\approx430$--950~K and $\Delta T_{\rm lat}\approx20$--110~K for 1~ps. Below the Debye temperature the lattice heat capacity is smaller than the Dulong--Petit value: for an illustrative Debye model with $\Theta_D=200$--250~K, a 100~fs pulse heats the lattice by 2.4--12.5~K from 150~K and by 4--23~K from 50~K. These estimates call for short pulses; continuous-wave driving would require a separate steady-state heat-flow analysis. The working window is $\hbar\omega\approx0.1$--0.25~eV with pulses of several optical cycles (one cycle lasts 41, 21 and 17~fs at 0.1, 0.2 and 0.25~eV, so five cycles take 207, 103 and 83~fs), about 100--200~fs at 0.2~eV ($\omega\tau\approx7.6$; $\approx3.8$ at 0.1~eV, where $P$ is 6\% below its $\omega\tau\gg1$ value). These estimates assume that the absorbed energy thermalizes among the electrons within the pulse and neglect transfer to the lattice during it; they constrain the deposited energy, not the momentum-resolved distribution. Main-text Fig.~4(a) uses the 50~K base.

\subsection*{S12.2. Charge-density wave}
Monolayer \NbSe{} develops a $3\times3$ charge-density wave below $T_{\rm CDW}\approx145$~K~\cite{S:Xi2015}, with a gap of about 4~meV at $\EF$~\cite{S:Ugeda2016}, which we neglect, as in Ref.~\cite{S:Wicaksono2026}. Our predictions therefore describe the unreconstructed normal state and apply most directly to samples held above $T_{\rm CDW}$, a reported value that may differ for supported or doped monolayers. From a 150~K base, the pulses that realize $\alpha=0.01$ and $0.02\angi$ heat the electrons to $T_e\approx212$ and $\approx335$~K ($\approx298$ and $\approx536$~K for the bare heat capacity; see above); the light-induced Hall conductivity at $\EF$ is $(-1.91,\allowbreak\,-1.93,\allowbreak\,-1.99,\allowbreak\,-2.00,\allowbreak\,-2.13)\times10^{-3}\eh$ at 50, 150, 300, 335 and 540~K (Sec.~S9), and the BCD peak (maximum over $\mu$) is $4.28\times10^{-2}$~\AA{} at 150~K and $(3.20,\allowbreak\,2.55)\times10^{-2}$~\AA{} at 335 and 540~K [$(3.15,\allowbreak\,2.50)\times10^{-2}$~\AA{} at the fixed density of the 150~K peak]. Below $T_{\rm CDW}$, the estimated lattice heating (4--23~K from 50~K) does not determine the transient dynamics of the electronically coupled CDW; a reconstructed-band, nonequilibrium analysis would be needed. Since the light-induced saddle splitting is only a few meV, a reconstruction near M could substantially modify the BCD peak; its effect on the Hall valve remains to be assessed. Superconductivity is irrelevant at these temperatures.

\subsection*{S12.3. Device estimates and competing signals}
We use the sheet conductance measured in monolayer \NbSe{} near $T_c$, $\sigma_{xx}\approx5$~mS, as an estimate at the driven electron temperatures (${\approx}200~\Omega$ per square)~\cite{S:Xi2016}, with residual resistance ratio ${\approx}10$, mean free path ${\approx}15$~nm, carrier density ${\approx}8.8\times10^{14}$~cm$^{-2}$ and $\tau\approx20$--150~fs (our estimate). For the valve at 0.4\MVcm{} (leading order, $|\slight(\EF)|=1.54\times10^{-3}\eh$), the Hall angle is $\sigma_{xy}/\sigma_{xx}=1.2\times10^{-5}$ and the transient Hall current is $I_H\approx\sigma_{xy}V\approx6$~nA at a dc bias of 0.1~V, i.e., ${\approx}0.004$ electrons per 100~fs pulse. This is the transverse current collected in a square geometry ($L=W$) with short-circuited Hall contacts; an open-circuit Hall bar instead measures a Hall voltage $\approx(\sigma_{xy}/\sigma_{xx})V$. The average current is suppressed by the pulse duty cycle $t_pf_{\rm rep}$, motivating time-resolved on-chip detection; the feasibility of dc averaging depends on the repetition rate and noise floor. For the light-induced BCD, the nonlinear Hall current $I^{(2)}=(e^3\tau/2\hbar^2)\,|D|\,(V/L)^2W$ for a $5~\mu$m device ($L=W$) at a bias amplitude of 0.1~V, i.e., the amplitude of the second-harmonic current (the rectified dc part has the same magnitude; Sec.~S2), is 40--310~pA at the hole-doped peak and 1--10~pA at $\EF$ for $\tau=20$--150~fs; being quadratic in the bias, it can be isolated as the second harmonic of a low-frequency bias. For circularly polarized light at normal incidence with $\hat{\bm m}\parallel\hat{\bm z}$, the ideal structure keeps $C_3$, which forbids in-plane dc photocurrents odd in $\eta$, so circular photogalvanic and photon-drag currents cannot mimic the helicity-odd Hall signal. Linear light, used for the BCD, breaks $C_3$; there intrinsic linear photogalvanic (shift) currents require real interband transitions, whose dominant onset in the model is 0.60~eV (0.42~eV at the hole-doped BCD peak; Sec.~S5), and any bias-independent photocurrent, including extrinsic ones, drops out of the second-harmonic detection. In a real device, contacts and substrate can lower the symmetry, so helicity and bias reversal should be combined to isolate the Hall signal. Possible probes are ultrafast on-chip transport~\cite{S:McIver2020} and circular-dichroism time-resolved photoemission, which maps Floquet Berry curvature in momentum space~\cite{S:Kim2026}.

% ---- references of the Supplemental Material (supplement.bbl, keys prefixed S:);
% labels set inside it (revtex: LastBibItem) get the prefix SM to stay distinct from the main reference list
\begingroup\let\SMoriglabel\label\def\label#1{\SMoriglabel{SM#1}}%
\endgroup

\end{document}